\newif\iffullversion
\newif\ifanonymous

\ifdefined\isfullversion
\fullversiontrue
\fi

\ifdefined\isanonymous
\anonymoustrue
\fi

\newif\ifshaphered

\ifdefined\isshaphered
\shapheredtrue
\fi

\documentclass[conference]{IEEEtran}
\usepackage{graphicx}
\usepackage{color}
\usepackage{footnote}
\usepackage{multirow}
\usepackage{mathtools}
\usepackage{booktabs}
\usepackage{multirow}
\usepackage{placeins}
\usepackage{float}
\restylefloat{table}
\usepackage{siunitx}
\usepackage{array}
\usepackage{verbatim}
\usepackage{url}
\usepackage{breakurl}
\usepackage[breaklinks]{hyperref}
\usepackage{longtable}

\usepackage{lipsum}
\usepackage[pscoord]{eso-pic}
\newcommand*{\textlabel}[2]{%
  \edef\@currentlabel{#1}% Set target label
  \phantomsection% Correct hyper reference link
  #1\label{#2}% Print and store label
}

\ifCLASSINFOpdf
\else
\fi
\begin{document}
\title{Practical Attacks Against Privacy and Availability in 4G/LTE
  Mobile Communication Systems}

\ifanonymous
%% No author names 
\else
%\author{\IEEEauthorblockN{Michael Shell}
%\IEEEauthorblockA{Georgia Institute of Technology\\
%someemail@somedomain.com}
%\and
%\IEEEauthorblockN{Homer Simpson}
%\IEEEauthorblockA{Twentieth Century Fox\\
%homer@thesimpsons.com}
%\and
%\IEEEauthorblockN{James Kirk\\ and Montgomery Scott}
%\IEEEauthorblockA{Starfleet Academy\\
%someemail@somedomain.com}}



\author{\IEEEauthorblockN{Altaf Shaik\IEEEauthorrefmark{1},
Ravishankar Borgaonkar\IEEEauthorrefmark{2},
N. Asokan\IEEEauthorrefmark{3}, 
Valtteri Niemi\IEEEauthorrefmark{4} and
Jean-Pierre Seifert\IEEEauthorrefmark{1}}
\IEEEauthorblockA{\IEEEauthorrefmark{1}Technische Universit{\"a}t Berlin and Telekom Innovation Laboratories\\
 Email: {(altaf329, jpseifert)} @sec.t-labs.tu-berlin.de}
\IEEEauthorblockA{\IEEEauthorrefmark{2}Aalto University\\
Email: ravishankar.borgaonkar@aalto.fi}
\IEEEauthorblockA{\IEEEauthorrefmark{3}Aalto University and University of Helsinki\\
Email: asokan@acm.org}
\IEEEauthorblockA{\IEEEauthorrefmark{4}University of Helsinki \\ Email: valtteri.niemi@helsinki.fi}}
\fi %% anonymous


\IEEEoverridecommandlockouts
\makeatletter\def\@IEEEpubidpullup{9\baselineskip}\makeatother
\IEEEpubid{\parbox{\columnwidth}{Permission to freely reproduce all or part
    of this paper for noncommercial purposes is granted provided that
    copies bear this notice and the full citation on the first
    page. Reproduction for commercial purposes is strictly prohibited
    without the prior written consent of the Internet Society, the
    first-named author (for reproduction of an entire paper only), and
    the author's employer if the paper was prepared within the scope
    of employment.  \\
    NDSS '16, 21-24 February 2016, San Diego, CA, USA\\
    Copyright 2016 Internet Society, ISBN 1-891562-41-X\\
    http://dx.doi.org/10.14722/ndss.2016.23236
}
\hspace{\columnsep}\makebox[\columnwidth]{}}

\maketitle

\ifshaphered
\newcommand{\on}[2]{{\textcolor{red}{#1}} \textcolor{blue}{#2}}
\input{shepherd-changelog}
\else
\newcommand{\on}[2]{#2}
\fi

%\lipsum[1-5]%
%\placetextbox{0.5}{1}{ \huge \color{red} \texttt{* Confidential - For internal review only *}}%

\begin{abstract}
%\boldmath
% Mobile devices have become a necessity in a human life, thanks to availability and bandwidth of LTE network services everywhere. The LTE communication protocols promise several appropriate levels of subscriber piracy for location and availability of network services all the time. In this work, we analyze access network security protocols of LTE networks. 
%  We expose three novel threats to the subscriber privacy in LTE networks, which make it possible to identify in 2 sq. km area of a city and potentially precisely locate subscribers with their GPS coordinates.
% In addition, we uncover three new threats to the availability of LTE networks to subscribers, which make it possible to persistently deny LTE or all types of mobile services to subscribers. We demonstrate feasibility of threats on commercially available LTE mobile phones using low cost implementation framework. We discuss deficiencies in LTE security architecture with respect to security and availability related to threats exploration. Finally we propose countermeasures and conclude that. 

Mobile communication systems are now an essential part of life
throughout the world. Fourth generation ``Long Term Evolution'' (LTE)
mobile communication networks are being deployed. The LTE
suite of specifications is considered to be significantly better than
its predecessors not only in terms of functionality but also with
respect to security and privacy for subscribers. We carefully analyzed LTE access network protocol specifications and
uncovered several vulnerabilities.  Using commercial LTE mobile
devices in real LTE networks, we demonstrate inexpensive, and
practical attacks exploiting these vulnerabilities. Our first
class of attacks consists of three different ways of making an LTE
device leak its location: \on {A semi-passive attacker can locate an LTE
device within a 2 $km^2$ area within a city}{\hypertarget{r4-2}{In our experiments, a semi-passive attacker can locate an LTE
device within a 2 $km^2$ area in a city}} whereas an active attacker
can precisely locate an LTE device using GPS co-ordinates or
trilateration via cell-tower signal strength information. Our second
class of attacks can persistently deny some or all services to a
target LTE device.  
\on {To the best of our knowledge, our work constitutes the first publicly reported
practical attacks against LTE.} {\hypertarget{r2-1}{To the best of our knowledge, our work constitutes
the  \textit{first publicly reported practical} attacks against LTE
access network protocols.}}

We present several countermeasures to resist our specific attacks. 
%also discuss possible trade-offs
%, between security and other criteria
%like availability and performance, 
\on {We that may explain why these vulnerabilities exist and recommend that
safety margins introduced into future specifications to address such
trade-offs should incorporate greater agility to accommodate subsequent
changes in the trade-off equilibrium.} {\hypertarget{r2-5}{%We also discuss why these vulnerabilities exist: e.g., justifications for trade-offs made when LTE specifications were developed are no longer valid. 
We also discuss possible trade-off considerations that may explain why these vulnerabilities exist. We argue that justification for these trade-offs may no longer valid. We recommend that safety margins introduced into future specifications to address such trade-offs should incorporate greater agility to accommodate subsequent changes in the trade-off equilibrium.}}

%\on{To the best of our knowledge, our work constitutes the  \textit{first publicly reported practical} attacks against LTE access network protocols.} {\hypertarget{com1}{To the best of our knowledge,}attacks against LTE access}.

\end{abstract}


% \fi
%%\IEEEpeerreviewmaketitle

 \section{Introduction}
\label{sec:intro}

During the past two decades, mobile devices such as smartphones have
become ubiquitous. The reach of mobile communication systems, starting
from the second generation Global System for Mobile Communications (2G/GSM) and the third generation Universal Mobile Telecommunication
Systems (3G/UMTS), has extended to every corner in the world. Mobile
communication is an important cornerstone in the lives of the vast
majority of people and societies on this planet. The latest generation
in this evolution, the fourth generation ``Long Term Evolution'' (4G/LTE) systems are being deployed widely. By the end of 2015
the worldwide LTE subscriber base is expected to be around 1.37
billion~\cite{ABIreport}.

Early 2G systems were known to have several vulnerabilities. For
example, lack of mutual authentication between mobile users
and the network implied that it was possible for an attacker to set up
fake base stations and convince legitimate mobile devices to
connect to it. In order to minimize exposure of user identifiers
(known as International Mobile Subscriber Identifier or IMSI) in
over-the-air signaling messages, 2G systems introduced the use of
temporary mobile subscriber identifiers. However, in the
absence of mutual authentication, fake base stations were used as
``IMSI catchers'' to harvest IMSIs and to track movements
of users.

The evolution of these mobile communication systems specified by 3GPP
(Third Generation Partnership Project) have not only incorporated
improvements in functionality but also strengthened security. 3G
specifications introduced mutual authentication and the use of
stronger and well-analyzed cryptographic algorithms. LTE
specifications further strengthened signaling protocols by requiring
authentication and encryption (referred to as ``ciphering'' in 3GPP
terminology) in more situations than was previously
required. Consequently, there is a general belief that LTE
specifications provide strong privacy and availability guarantees to
mobile users. Previously known attacks, such as the ability to track
user movement were thought to be
difficult in LTE.

\on{In this paper, we demonstrate the first practical attacks
against LTE devices.} {\hypertarget{r4-5}{In this paper, we demonstrate the \textit{first practical} attacks
against LTE access network protocols.}} Our attacks are based on
vulnerabilities we discovered during a careful analysis of LTE access
network protocol specifications. They fall into two classes: location
leaks and denial of service. In the first class, we describe three
different attacks that can force an LTE device (User Equipment or UE
in 3GPP terminology) into revealing its location. \on{The first two allow
a passive or semi-passive attacker to localize the target user within
about a 2 $km^2$ area in an urban setting which is a much finer
granularity than previously reported location leak
attacks~\cite{{fookune2010locationgsm}} against 2G devices.} {\hypertarget{r1-1} {The first two allow
a passive or semi-passive attacker to localize the target user within
about a 2 $km^2$ area in an urban setting which is a much finer
granularity than previously reported location leak
attacks~\cite{{fookune2010locationgsm}} against 2G devices, while still using similar techniques.}} Notably, we show how popular social network messaging
applications (e.g., Facebook messenger~\cite{fmessg} and WhatsApp~\cite{whatsapp}) can be used in such attacks. Our third attack allows an
active attacker exploiting vulnerabilities in the specification and
implementation of LTE Radio Resource Control (RRC)
protocol~\cite{36.331} to accurately pinpoint the target user via GPS
co-ordinates or trilateration using base station signal strengths as
observed by that UE. \textit{We believe that all LTE devices in the market are
  vulnerable to this attack}.
 
In the second class, we describe three further attacks where an active
attacker can cause persistent denial of service against a target
UE. In the first, the target UE will be forced into using 2G or
3G networks rather than LTE networks, which can then make it possible
to mount 2G/3G-specific attacks against that UE. In the second, the target UE will be denied access to all networks. In the
last attack, the attacker can selectively limit a UE only to some types of
services (e.g., no voice calls). The attacks are persistent and silent:
\textit{devices require explicit user action (such as rebooting the
  device) to recover}.

We have implemented all our attacks (except one) and confirmed their
effectiveness using commercial LTE devices from several vendors and
real LTE networks of several carriers. The equipment needed for
the attacks is inexpensive and readily available. We reported our
attacks to the manufacturers and carriers concerned as well as to the
standardization body (3GPP). %At the time of writing, remedial actions are under way.
Remedial actions are under way while writing.
%% Later we can provide more specific information about the remedial actions.

Specification of a large system like LTE is a complex endeavor
involving many trade-offs among conflicting requirements. Rather than
merely report on LTE vulnerabilities and attacks, we also discuss
possible considerations that may have led to the vulnerabilities in
the first place. Based on this we suggest some general guidelines for
future standardization as well as specific fixes for our attacks.

\begin{itemize}

	\item {Fine-grained location leaks:} New passive and active techniques to link users' real identities to LTE temporary identities assigned to them and to \textbf{track user locations and movements to much higher levels of granularity} than was previously thought possible. (Section~\ref{sec:location-attacks}) % This attack is applicable to \textbf{every LTE Advanced mobile device}.
	%We discovered vulnerability in LTE 3GPP standard specification that allows precise tracking of subscribers. The vulnerability affects to every LTE Advance mobile phone available in the market.
	%Further we identify network configuration issues responsible for subscriber's location leak in three European operator's LTE networks. 

	\item {Denial-of-Service (DoS) Attacks:} New active \textbf{DoS attacks that can silently and persistently downgrade LTE devices} by preventing their access to LTE networks (limiting them to less secure 2G/3G networks or denying network access altogether) or limiting them to a subset of LTE services. (Section~\ref{sec:dosattacks})
%against LTE subscribers. As a consequence, LTE subscribers can be downgraded forcefully to use attack-prone GSM or 3G networks and denied use of all mobile networking services silently. 
		
	\item {Implementation \& Evaluation:} \textbf{Inexpensive software and hardware framework} to implement the attacks based on \texttt{srsLTE}, \texttt{OpenLTE}, and Universal Software Radio Peripherals (USRP) (Section~\ref{sec:experiment}), and evaluation of the attacks using commercially available LTE phones in real networks. (Sections \ref{sec:location-attacks}--\ref{sec:feasi}) %Discovered vulnerabilities via software-framework and network configuration issues responsible for location leaks are notified to baseband vendors and to network operators respectively.  

	\item {Security Analysis:} Discussion outlining \textbf{possible underlying reasons for the vulnerabilities}, including perceived or actual trade-offs between security/privacy and other criteria like availability, performance and functionality, as well as recommending fixes. (Section~\ref{sec:securityanalysis}).
%We highlight the design concerns of the LTE security protocols related to these attack exploration and experimentation on different baseband manufacturers and deployed LTE networks. 

\end{itemize}

 \section{Overview of LTE Architecture }
\label{sec:background}

%In this section, we briefly describe LTE infrastructure and paging mechanisms in order to assist readers to understand attacks presented in this paper. In addition, we give an overview of security architecture and main features that LTE promises. 

We briefly describe LTE infrastructure as well as security
and paging mechanisms to assist readers in understanding the vulnerabilities and attacks we present in this paper.

\subsection{LTE infrastructure}

%We consider simplified LTE architecture involving components required to setup access network security protocols between the base station and mobile phones. We hide other details of the architecture which are uninteresting for the purpose of demonstrating our attacks. Figure~\ref{fig:ltearchi} depicts these components and described as below.  

\begin{figure}[htbp]
  \begin{center}
  \includegraphics[width=\linewidth]{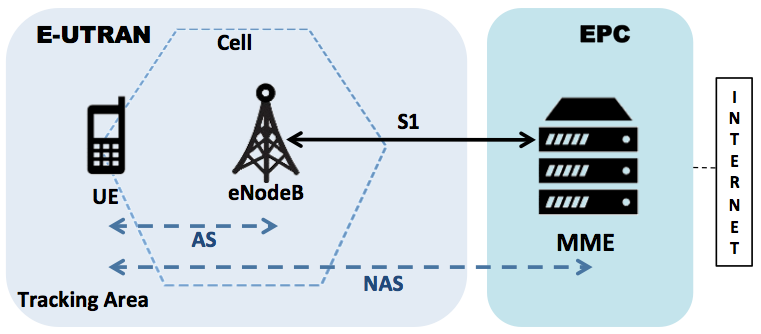}
  \caption{LTE system architecture}
  \label{fig:ltearchi}
  \end{center}
\end{figure}

We consider a simplified LTE architecture involving components
required to set up access network protocols between a base station and
mobile devices. We hide other details of the architecture which are
not relevant from the point of view of understanding our
attacks. Figure~\ref{fig:ltearchi} depicts this simplified
architecture which contains three main components: User Equipment
(UE), Evolved Universal Terrestrial Radio Access Network (E-UTRAN),
and Evolved Packet Core (EPC). All three components are collectively
referred to as Evolved Packet System (EPS) according to 3GPP
terminology. In the interest of simplicity, throughout this paper we
refer to the whole system as LTE.  The three components are described
below (A list of common acronyms related to LTE appear in the full version of this paper~\cite{fullpaper}).
%Table~\ref{tab:abbrev}).

%The basic LTE system architecture contains three main components, namely User Equipment (UE), Evolved Universal Terrestrial Radio Access Network (E-UTRAN), and Evolved Packet Core (EPC). All three components are collectively referred as Evolved Packet System (EPS) according to 3GPP terminology. However, throughout this paper we refer the whole system as LTE. Figure~\ref{fig:ltearchi} depicts the LTE system architecture and the components which are described below.

%\textbf{User Equipment:} UE refers to the actual communication device which can be a smartphone or M2M device~\cite{m2m}. A UE requires a UICC (Universal Integrated Circuit Card)~\cite{31.101}, colloquially known as SIM card in order to connect to the network and access LTE services. It runs an application known as USIM (Universal Subscriber Identity Module)~\cite{31.102}, which stores user specific data and is uniquely identified by the IMSI (International Mobile Subscriber Identity)~\cite{23.003}. The USIM is used to identify and authenticate subscribers and to derive security keys for protecting their communication over the radio interface.
\noindent\textbf{User Equipment:} UE refers to the actual communication device which can be, for example, a smartphone \iffullversion or a machine-to-machine (M2M) communication device~\cite{m2m}\fi. A UE contains a USIM (Universal Subscriber Identity Module)\cite{31.102}, which represents the IMSI and stores the corresponding authentication credentials~\cite{23.003}. This IMSI is used to identify an LTE user (generally referred to as ``subscriber'' in 3GPP terminology) uniquely. The USIM participates in LTE subscriber authentication protocol and generates cryptographic keys that form the
basis for the key hierarchy subsequently used to protect signaling and user data communication between the UE and base stations over the radio interface.
%generates cryptographic keys subsequently used to protect signaling and user data communication between the UE and base stations over the radio interface.

\noindent\textbf{E-UTRAN:} E-UTRAN consists of base stations. It manages the radio communication with the UE and facilitates communication between the UE and EPC. In LTE, a base station is technically referred as ``evolved NodeB (eNodeB)''. The eNodeB uses a set of access network protocols, called Access Stratum (AS) for exchanging signaling messages with its UEs. These AS messages include Radio Resource control (RRC) protocol messages. Other functions of eNodeB include paging UEs, over-the-air security, physical layer data connectivity, and handovers. Each eNodeB is connected to the EPC through an interface named S1.% and can be connected to nearby eNodeBs.

%E-UTRAN solely consists of evolved NodeB (eNodeB), and manages the radio communications with the UE and serves the relay in communication between the UE and EPC. The eNodeB is a base station whose main functions are exchanging signalling messages and data to all its UEs both on downlink and uplink. Radio data connection, paging, over-the-air security, handovers are some of the important functions handled by the eNodeB. The set of protocols running between the eNodeB and UE are referred as Access Stratum (AS) messages. AS messages are also referred as RRC (Radio Resource control) messages. Each eNodeB is connected to the EPC through a S1 interface and can be connected to other nearby eNodeBs. 

\noindent\textbf{MME in EPC:}
EPC provides core network functionalities by a new all-IP mobile core network designed for LTE systems. It consists of several new elements as defined in~\cite{23.002}. However, for our work we need to describe only the Mobility Management Entity (MME) in detail. MME is responsible for authenticating and allocating resources (data connectivity) to UEs when they connect to the network. Other important functions of MME involve security (setting up integrity and encryption for signaling)~\cite{33.401} and tracking UE's location at a macro level. The set of protocols run between UE and MME are referred as Non-Access Stratum (NAS).

Now, we explain how the system components presented above can be deployed in a geographical region (e.g., in a city) by mobile network carriers (more commonly referred to as ``operators'' in 3GPP terminology) to provide LTE services. A service area of a mobile operator is geographically divided into several regions known as Tracking Areas (TAs). TAs are similar to Location Areas in GSM networks and are managed by the MME. Further, a TA contains a group of ``cells''\footnote{In LTE, coverage area of an eNodeB is divided into several sectors known as cells.} each of which is controlled by an eNodeB.  
The eNodeB broadcasts operator-specific information such as Tracking Area Code (TAC), Mobile Country Code (MCC), Mobile Network Code (MNC), and cell ID via System Information Block (SIB) messages~\cite{36.331}. This allows UEs to identify their serving network operator, and initiate a connection to the network. A UE attaches to the network by initiating the $Attach$ procedure~\cite{24.301}. Upon successful acceptance the UE receives access to services based on its subscription. The UE uses the $Tracking Area Update (TAU)$ procedure to inform the network about its mobility in the serving area~\cite{24.301}.

\subsection{Security in LTE }

As IMSI is a permanent identifier of a subscriber, LTE specifications try to minimize its transmission in over-the-air radio communication for security and privacy reasons. Instead, a Globally Unique Temporary Identifier (GUTI)~\cite{23.003} is used to identify subscribers during radio communication. It is assigned to UEs during $Attach$ and may be periodically changed to provide temporal unlinkability of traffic to/from the same UE.
An Authentication and Key Agreement (AKA) protocol is used for mutual authentication between UE and the network and to agree on session keys that provide integrity and confidentiality protection for subsequent NAS and AS messages~\cite{33.401}. 
Both NAS and AS security are collectively referred as EPS security.
It is established between a UE and a serving network domain (eNodeB and MME) during EMM (EPS Mobility Management) procedures~\cite{24.301} and includes agreeing on session keys, preferred cryptographic algorithms, and other values as defined in~\cite{33.401}. 
%In this paper, we analyze protocol security issues on both NAS and AS layers involving the UE, eNodeB, and the MME entities.

%Since IMSI is a permanent identity of subscribers,  LTE avoids its transmission over the air-interface for security and privacy reasons. Instead, GUTI (Globally Unique Temporary Identifier)~\cite{23.003} is used to identify subscribers during the radio communication. It is assigned to UEs during the attach process~\cite{24.301} and may be periodically changed to prevent their tracking by unauthorized parties.

%LTE defines security procedures to protect NAS and AS messages exchanged over the radio interface~\cite{33.401}. In specific, these procedures utilize an Authentication and Key Agreement (AKA) mechanism to provide integrity and confidentiality protection. Both NAS and AS security are collectively referred as EPS security context. The security context is established between the UE and a serving network domain (eNodeB and MME) during EMM (EPS mobility Management) procedures~\cite{24.301} and includes security keys, preferred cryptographic algorithms, and other values as defined in~\cite{33.401}. In this paper, we analyze protocol security issues on both NAS and AS layers involving the UE, eNodeB, and the MME entities.

\subsection{Paging in LTE}
\label{sec:paging}

\begin{figure}[!t]
  \begin{center}
   \includegraphics[width=3.5in]{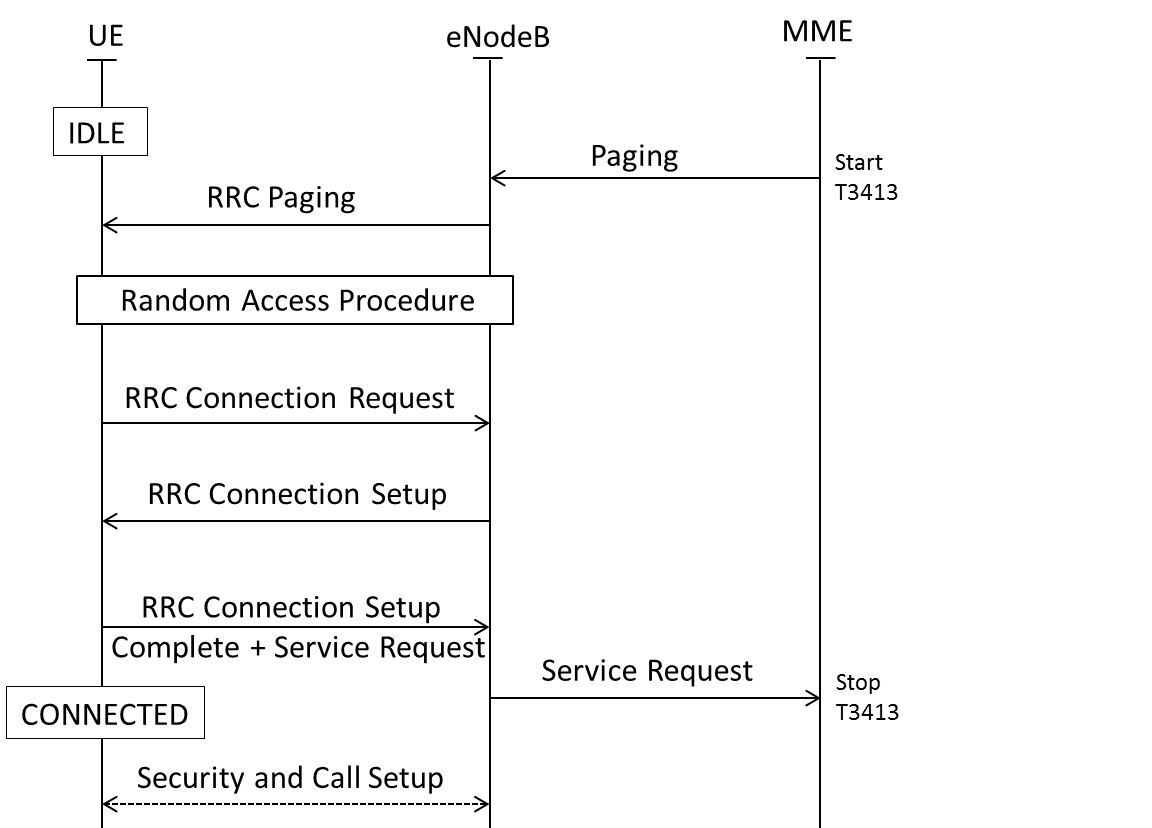}
   \caption{Paging in LTE}
   \label{fig:LTEPagingProcedure}
   \end{center}
 \end{figure}

Paging refers to the process used when MME needs to locate a UE in a particular area and deliver a network service, such as incoming calls.
Since MME may not know the exact eNodeB to which UE is connected, it generates a paging message and forwards to all eNodeBs in a TA. Simultaneously, MME starts a paging timer (T3413) and expects a response from UE before this timer expires.  Thus, all eNodeBs present in the paged TA broadcast a RRC paging message to locate the UE. Paging messages contain identities of UEs such as  S-TMSI(s) or IMSI(s). S-TMSI is a temporary identifier (SAE-Temporary Mobile Subscriber Identity). It is part of a GUTI. For the sake of simplicity, we consistently use the term GUTI throughout the rest of this paper even when referring to S-TMSI. Figure~\ref{fig:LTEPagingProcedure} highlights LTE paging procedure, described in detail in the relevant LTE specifications ~\cite{24.301,36.304,36.300}.

%The paging procedure for LTE services is defined in these~\cite{24.301,36.304,36.300} 3GPP specifications. It is applicable to UEs that are registered on the network and are in IDLE state. When UE has an incoming call or data the MME initiates a paging procedure. Since the MME does not know the exact eNodeB to which UE is connected to, it generates a paging message and forwards to all other eNodeBs in a TA (or a TA list).  Simultaneously the MME starts a paging timer (T3413) and expects a response from UE before the expiry of this timer. All base stations (eNodeBs) present in the paged TA prepare a RRC paging message and broadcast it during  appropriate paging occasion. Paging occasion refer to the moment where the UE can expect a paging message from nearby eNodeBs. Paging message contain S-TMSI (s) or IMSI (s) of the UE (s) being paged where S-TMSI is a part of GUTI. Note that GUTI can be constructed by knowing the S-TMSI and therefore in this paper, we refer GUTI as a temporary identity instead of S-TMSI. Figure~\ref{fig:LTEPagingProcedure} highlights this LTE paging procedure.

The UE in IDLE state\footnote{In IDLE state, the UE has no active connections with any eNodeB.} decodes RRC paging messages. If it detects its IMSI then it initiates a new $Attach$ procedure to receive a GUTI as defined in~\cite{24.301}. If UE detects its GUTI, it acquires a radio channel through the \emph{``Random Access Procedure"}~\cite{36.300} for requesting a RRC connection from the eNodeB. \emph{``RRC Connection Setup"} involves the configuration of radio resources for exchanging signaling messages. Upon receiving this setup message, the UE completes a three way RRC handshake procedure by sending a \emph{``RRC Connection Setup Complete"} message along with a \emph{``Service Request"} message. At this point UE leaves IDLE state and enters into CONNECTED state\footnote{CONNECTED means the UE has an active connection with an eNodeB.}. The eNodeB forwards the service request message to MME, which in turn stops the paging timer. Further, eNodeB establishes a security context and proceeds to deliver network services to UE.

In LTE, the paging procedure is improved to reduce signaling load and locate the UE faster using a technique called Smart Paging~\cite{melih,alcatel,nokia}. It is compliant with LTE specifications and consists of directing paging messages selectively via the eNodeB (cell) where the UE was last seen. If no response is received, paging is repeated in the entire TA. In our experiments (Section~\ref{sec:initialmesurements}) to study LTE paging procedures in a major city, we observed that several network operators and vendors have implemented smart paging.

%As LTE is an all-IP network, the core signaling volumes are significantly higher than in existing 2G/3G networks. The rise in core signaling can also be attributed to an overall increase in network usage by LTE subscribers  due to smartphones and applications. Hence, it is not efficient to page a complete TA to find a particular UE which incurs signalling load on network components~\cite{alcatel}. In our experiments (Section~\ref{sec:initialmesurements}) to study LTE paging procedures in a major city , we observed that operators and vendors have implemented smart paging techniques into their core network. According to Alcatel-Lucent~\cite{alcatel}, a smart paging technique allows the MME to page UEs on the last seen cell (eNodeB) instead of a complete TA. If the MME does not receive a paging response back before the expiry of associated paging timer, it sends paging messages to the complete TA. As a result, signalling load on the core network is reduced. Several vendors~\cite{melih}~\cite{alcatel}~\cite{nokia} are implementing optimized TA and smart paging algorithms into their MME products that can enable networks to find the UE (subscriber) faster. Moreover, these methods are fully compliant with 3GPP standards~\cite{alcatel}.

 \section{Adversary Model}
\label{sec:advsersarymodel}

In this section, we describe the adversary model for our attacks. The primary goals of the adversary against LTE subscribers are: a) learn the precise location of a subscriber in a given geographical area b) deny network services (both mobile-terminated and mobile-originated) to a subscriber, and c) force subscribers to use less secure GSM or 3G networks thereby exposing them to various attacks such as IMSI catchers~\cite{strobel2007imsi}. \on{} {\hypertarget{r4-3}{We assume that the adversary is in the same geographical area as the victim.}} The adversary model is divided into three attack modes as described below.

\subsection*{Passive}
A passive adversary is able to silently sniff LTE over-the-air (radio) broadcast channels. To achieve this, he/she has access to a hardware device (for example Universal Software Radio Peripheral (USRP)) and associated software needed to observe and decode radio broadcast signaling messages.

\subsection*{Semi-Passive}
A semi-passive adversary is, in addition to passive monitoring, able to trigger signaling messages to subscribers using interfaces and actions that are \emph{legitimately available} in LTE or in higher layer systems. For example, a semi-passive adversary can trigger paging messages to subscribers by sending a message via a social network or initiating a call. The adversary is assumed to be aware of social identities of subscribers. For example, these identities can be a Facebook profile or a mobile phone number of the subscriber. A semi-passive adversary is analogous to the `honest-but-curious' or `semi-honest' adversary model used for cryptographic protocols~\cite{Goldreich2004}.

\subsection*{Active}
The active adversary can set up and operate a rogue eNodeB to establish malicious communication with UEs. Capabilities required for active attacks include knowledge of LTE specifications and hardware (USRP) that can be used for impersonating subscriber's serving operator network, and injecting malicious packets to UEs. An active adversary is analogous to the `malicious' adversary model in cryptographic protocols~\cite{Goldreich2004}.

 \section{Experimental Setup}
 \label{sec:experiment}

%Unlike in GSM, there are only a few LTE specific open-source software and hardware modules available for conducting experiments on LTE networks and UEs. 
%In this section, we describe the hardware, software and configurations used to perform attacks presented in this paper. We use the same hardware for both type of attacks but the implementation differs. Figure~\ref{fig:experimentalsetup} depicts the experimental setup.

\begin{figure}
  \includegraphics[width=\linewidth]{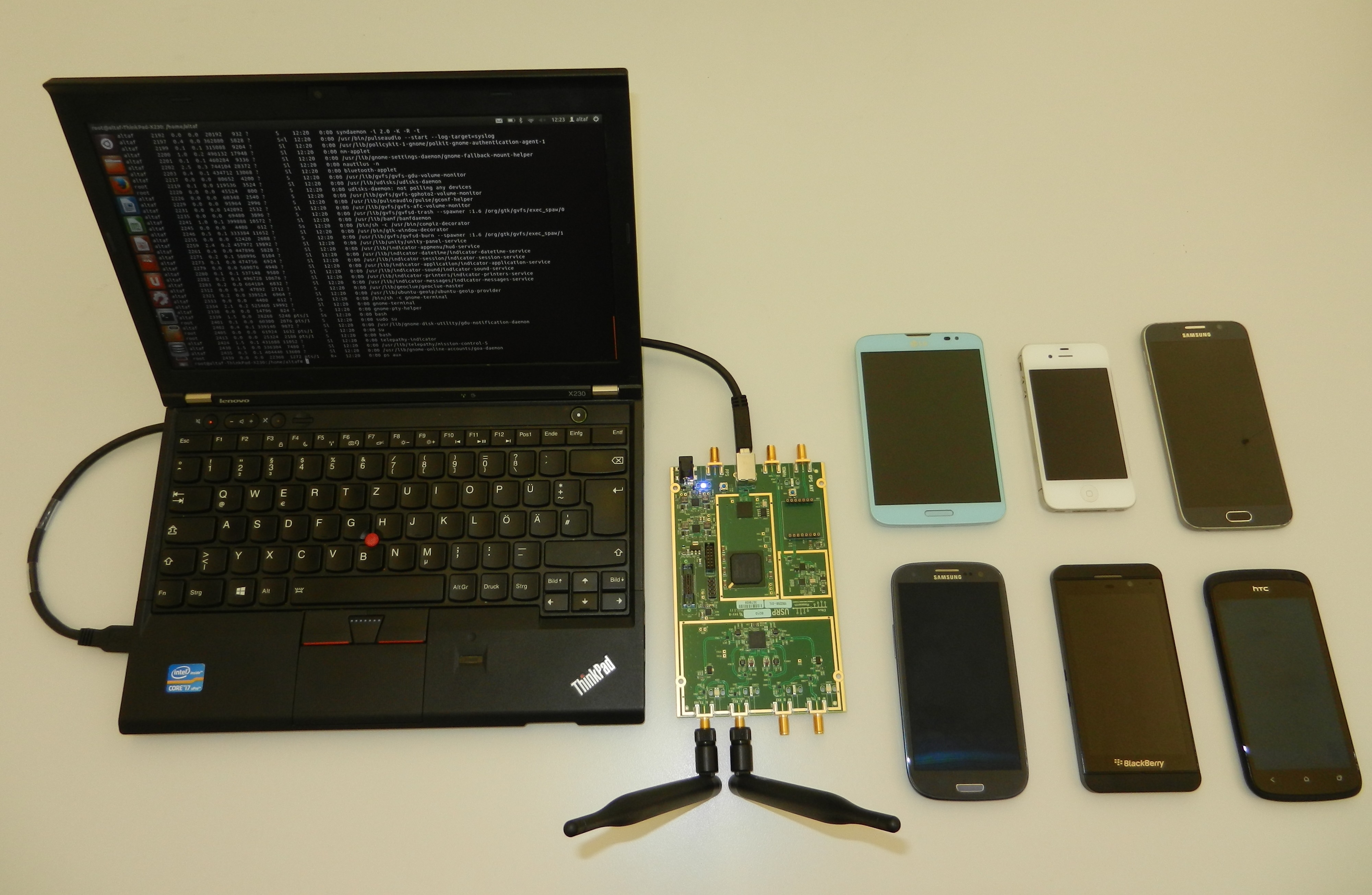}
  \caption{Experimental setup}
  \label{fig:experimentalsetup}
\end{figure}

%In traditional telecommunication systems, software and hardware are typically proprietary (closed source) and expensive. 
Software and hardware used in major telecommunication systems have traditionally been proprietary (closed source) and expensive.
However recently open source telephony software and low-cost hardware modules have started to emerge. In this section, we explain our experimental setup built using low cost off-the-shelf components and requiring only elementary programming skills with knowledge of LTE specifications. %Specifically in this section, we describe the hardware, software, and configurations used to build an LTE network infrastructure to perform attacks presented in this paper. 
Figure~\ref{fig:experimentalsetup} depicts the experimental setup.

\subsection*{Hardware}

Hardware components for eNodeB, MME, and UE are needed to build our experimental LTE network. On the network side, we used a USRP B210 device~\cite{usrp} connected to a host laptop (Intel i7 processor \& Ubuntu 14.04 OS), acting as an eNodeB. USRP is a software-defined radio peripheral that can be connected to a host computer, to be used by host-based software to transmit/receive data over the air. Even though we utilized USRP B210 which costs around one thousand euros, passive attacks can also be realized practically with more cheaply available radio hardware. For example, RTL-SDR~\cite{rtl-sdr} dongles which cost around 15 euros can be leveraged to passively listen over the LTE air-interface.
\iffullversion 
However RTL-SDR devices are not as stable as USRP due to hardware limitations.
\fi On the UE side, we selected popular LTE-capable mobile phones available in the market. These devices incorporate LTE implementations from four major LTE baseband vendors who collectively account for the vast majority of deployed LTE-capable UEs.

%TODO Ravi - add one sentecen about market share here . Note that these four vendors have majority market share 

%The hardware consists of a host laptop / Personal Computer (Intel i7 processor \& Ubuntu 14.04 OS), mobile phones with LTE support, and a USRP. USRPs are software-defined radios that can be connected to a host computer, to be used by host-based software to transmit/receive data over the air. Specifically, we used the low-cost USRP B210 device~\cite{usrp} for all attacks. It supports LTE frequency bands and is connected to the host via a high-speed USB3 link. Even though we utilized a USRP which costs around one thousand euros, passive attacks can be realized practically with other cheaply available radio hardware. For e.g., RTL-SDR~\cite{rtl-sdr} dongles which cost around 15 euros can be leveraged to passively listen over LTE air-interface. However USRP devices are more stable than RTL-SDR due to hardware limitations. There are mainly four different LTE baseband modem providers: Qualcomm, MediaTek, Samsung, and Huawei. We used popular mobile phones available in the market supporting LTE baseband for demonstrating attacks and their impact.

\subsection{Passive and semi-passive attack setup}

%The research test-bed used in performing paging attacks described in~\cite{fookune2010locationgsm} are restricted to GSM networks due to the unavailability of any LTE baseband implementations at that time. Today, there are some partial LTE baseband implementations available as open source including \texttt{OpenLTE}~\cite{openlte} and \texttt{srsLTE}~\cite{liblte}, which enabled us to conduct real-time experiments on LTE networks. We modified these to demonstrate the feasibility of the attacks we discovered. 

The research test-bed used in performing paging attacks described in~\cite{fookune2010locationgsm} was restricted to GSM networks due to the unavailability of any LTE baseband implementations at that time. Today, there are some partial LTE baseband implementations available as open source including \texttt{OpenLTE}~\cite{openlte} and \texttt{srsLTE}~\cite{liblte}, which enabled us to conduct real-time experiments on LTE networks.

\subsection*{Implementation}

In order to sniff LTE broadcast channels, we utilized parts of \texttt{srsLTE}. It is a free library for software-defined radio mobile terminals and base stations. Currently, the project is developing a UE-side LTE baseband implementation. \texttt{srsLTE} uses Universal Hardware Device library to communicate with the USRP B210. Since all the passive sniffing is done in real-time, it is recommended to have a high-speed host (laptop) in order to handle the high (30.72 MHz) sampling rates without data loss and also to maintain constant sync with eNodeBs. In particular, we used the \texttt{pdsch-ue} application to scan a specified frequency and detect surrounding eNodeBs. It can listen and decode SIB messages broadcast by eNodeB. Further, we modified \texttt{pdsch-ue} to decode paging messages which are identified over-the-air with a Paging-Radio Network Temporary Identifier (P-RNTI). Upon its detection, GUTI(s) and/or IMSI(s) can be extracted out of paging messages.

In semi-passive attack mode, we use Facebook~\cite{facebook,fmessg} and WhatsApp~\cite{whatsapp} applications over the Internet, in addition to initiating communication with targets via silent text messages or phone calls.

\subsection{Active attack setup}
\label{buildenodeb}

%The attack requires an eNodeB that can operate signalling messages on the LTE air-interface. In order to mount successful active attacks, the attacker requires a rogue eNodeB to inject unintentional traffic to UEs. The active attack is performed on an LTE (also 2G and 3G) capable UE which is currently registered to a commercial LTE network. The process of building such rogue eNodeB is described below. 
We built an eNodeB to mount successful active attacks against UEs registered with a real LTE network. In particular, our eNodeB impersonates a real network operator and forces UEs to attach to it. The process of building such rogue eNodeBs is described below.

\noindent\textbf{Building rogue eNodeB:} Generally, UE always scans for eNodeBs around it and prefers to attach to the eNodeB with the best signal power. Hence in IMSI catcher type of attacks~\cite{strobel2007imsi}, rogue eNodeBs are operated with higher power than surrounding eNodeBs. However, in LTE the functionality of the UE may be different in some situations. In particular, when a UE is very close to a serving eNodeB it does not scan surrounding eNodeBs. This allows UEs to save power. Hence to overcome this situation in our active attacks, we exploit another feature named \emph{`absolute priority based cell reselection'}, and introduced in the LTE release 8 specification~\cite{36.133}.

The principle of priority-based reselection is that UEs, in the IDLE state,  should periodically monitor and try to connect to eNodeBs operated with high priority frequencies~\cite{36.133}. Hence even if the UE is close to a real eNodeB, operating the rogue eNodeB on a frequency that has the highest reselection priority would force UEs to attach to it. These priorities are defined in SIB Type number 4, 5, 6, and 7 messages broadcast by the real eNodeB~\cite{36.331}. Using passive attack setup, we sniff these priorities and configure our eNodeB accordingly.

Further, the rogue eNodeB broadcasts MCC and MNC numbers identical to the network operator of targeted subscribers to impersonate the real network operator. Generally, when UE detects a new TA it initiates a \emph{``TAU Request"} to the eNodeB. In order to trigger such request messages, the rogue eNodeB operates on a TAC that is different from the real eNodeB.

\subsection*{Implementation}
%The attack is launched with the USRP B210 and a host laptop/PC together running an eNodeB application. This application is available from the open source project named OpenLTE. It is an open source implementation of 3GPP LTE specifications and encloses an eNodeB application. Although the eNodeB application cannot be compared to full-fledged commercial eNodeB, it has the capability to execute a complete LTE attach procedure. In addition, a few functionalities of the MME are implemented in the eNodeB application. Upon a successful completion of the attach procedure, the eNodeB can also handle mobile originated services, however, currently it lacks stability. We tested active attacks on UEs with USIMs from all three German operators.

%Further, we programmed the eNodeB application to include RRC and NAS protocol messages to demonstrate active attacks. In addition,
%we modified telephony protocol dissector~\cite{gsmtap} available in existing Wireshark application~\cite{Wireshark} to decode all the messages exchanged between the eNodeB and the UE. 
%These modifications are submitted to Wireshark project and are being merged into the mainstream application.

The active attack is launched using the USRP B210 and a host laptop which together are running \texttt{OpenLTE}. The \texttt{OpenLTE} is an open source implementation of LTE specifications and includes an \texttt{LTE\_Fdd\_enodeb} application. Although this application cannot be compared to a full-fledged commercial eNodeB, it has the capability to execute a complete LTE $Attach$ procedure. In addition, some functionality of the MME is implemented in \texttt{LTE\_Fdd\_enodeb}. Upon successful completion of $Attach$, \texttt{LTE\_Fdd\_enodeb} can also handle UE-originated services. However, currently it lacks stability. We tested active attacks on UEs with USIMs from three major national-level operators.

Further, we programmed \texttt{LTE\_Fdd\_enodeb} to include LTE RRC and NAS protocol messages to demonstrate active attacks. In addition, we modified the telephony protocol dissector~\cite{gsmtap} available in Wireshark~\cite{Wireshark} to decode all messages exchanged between the rogue eNodeB and UE.
These modifications are submitted to the Wireshark project and are being merged into the mainstream application.

\subsection{Ethical considerations}
Our work reveals vulnerabilities in LTE specifications which are already in use in every LTE-enabled UE worldwide. Further we also encountered several implementation issues in popular smartphones and LTE network configuration issues. Therefore we made an effort to responsibly disclose our work to the relevant standard bodies and affected parties. Our reports were acknowledged by all vendors and network operators we contacted. 
%We followed standard responsible disclosure processes that these parties already had in place.
%In the paper, we want to state that we followed responsible disclosure processes for vendors (who have such processes in place).
For those vendors who have a standard responsible disclosure process in place, we followed the process.
%We informed vulnerabilities discovered during our research to relevant LTE baseband vendors, 3GPP group, and network operators. These are acknowledged by them and we are following their responsible disclosure program. 

We carried out most of the active attacks in a Faraday cage~\cite{cage} to avoid affecting other UEs.
%As a caution, most of the attacks are carried out in a faraday cage.  
For attacks in real LTE networks, we took care not to interrupt normal service to other UEs in the testing zone. Initially, we determined GUTIs of our test UEs via passive attacks and fed them into our rogue eNodeB. We programmed our rogue eNodeB to accept  \emph{``TAU / Attach / Service Requests"} only from these specified GUTIs and to reject all requests from unknown UEs with the EMM reject cause number 12 \emph{``Tracking area not allowed"}~\cite{24.301}. Upon receipt of this message, all UEs other than our test UEs disconnect automatically from our rogue eNodeB.

 \section{Location Leak Attacks Over Air Interface}
\label{sec:location-attacks}

%In this section, we show how the approximate location of an LTE subscriber inside a metropolitan city can be inferred by applying a set of novel passive, semi-passive, and active attacks. In particular, first we track down the location of a subscriber to a cell level (2 sq. km area) using passive attacks and further determine the precise position using active attacks.  We first describe the background for the attacks by summarizing the features and aspects of LTE that are used in by the attacker. We then describe preliminary measurements used for realizing the attacks and new techniques for triggering subscriber paging. Finally we describe the attacks in detail.
In this section, we show how the approximate location of an LTE subscriber inside an urban area can be inferred by applying a set of novel passive, semi-passive, and active attacks. In particular, we track down the location of a subscriber to a cell level (e.g., 2 $ km ^ 2 $ area) using passive attacks (L1\footnote{For the sake of simplicity, we refer location leaks attacks as L1, L2, and L3 whereas DoS attacks as D1, D2, and D3 respectively.}) and further determine the precise position using active attacks (L3).  
We first describe the background for the attacks by summarizing the features and aspects of LTE that are used by the attacker. We then characterize preliminary measurements used for realizing the attacks and new techniques for triggering subscriber paging. Finally, we explain the attacks in detail.

\subsection{Attack background}
\label{network:issues}
We now describe network configuration issues, subscriber identity mapping technique, and observations about certain LTE network access protocols. We will later make use of all of these aspects in developing our attacks.
% which we will later make use of in developing our attacks.  %First, we discuss how passive and semi-passive attackers can exploit two network configuration issues. Then, we present a new method of mapping GUTIs to subscriber identities in LTE. Lastly, we uncover specification and implementation issues in LTE RRC protocol that can be exploited by an active attacker.
\subsection*{Network configuration issues}
In LTE, network operators deploy various methods to minimize signaling overhead introduced due to evolution of networks, devices, and smartphone applications~\cite{nokiablog}. We identify two such deployment techniques relevant to our discussion.

\noindent\textbf{Smart Paging:} In GSM, paging messages are sent to an entire location area. Thus it only allows the attacker to locate a subscriber within a large (e.g., 100 $km^2$) area~\cite{fookune2010locationgsm}. However, LTE paging is directed onto a small cell rather than to a large TA. Such Smart Paging allows an attacker to locate an LTE subscriber within a much smaller (e.g., 2 $km^2$) area which is a typical LTE cell size as observed in our experiments in a major city.

\noindent\textbf{GUTI persistence:} Generally a fresh GUTI is allocated in the following situations: (a) when MME is changed due to handover or load balancing, b) during TAU or $Attach$ procedure, and  c) when network issues NAS \emph{``GUTI reallocation command"}. However, network operators tend to not always change GUTI during the above procedures~\cite{lteload}\footnote{The reason for not changing GUTIs often is to avoid signaling storms in LTE network as described in~\cite{lteload}.}. This allows a passive attacker to track UEs based on their GUTIs.

%Generally a fresh GUTI is allocated in the following situations: (a) when MME is changed due to handover or load balancing, (b) when the GUTI timer (<name>) expires, or (c) during tracking area update or attach procedure. However, we observed that operators tend to not change GUTI sufficiently during the above procedures\footnote{The reason for not changing GUTIs often is to avoid signaling storm in LTE network as described in~\cite{lteload}.}. This allows a passive attacker to track UEs based on its GUTI.

\subsection*{Social identity to subscriber mapping} In previous work, 
%Public Switched Telephone Network call procedure
phone calls (originating from a landline phone)~\cite{fookune2010locationgsm} and silent Short Message Service (SMS)~\cite{sylvainCCC2010} techniques were used to page GSM subscribers thereby mapping TMSIs to their phone numbers. However these methods are not as effective anymore due to the availability of tools to detect such attacks~\cite{darshak,snoop}. We now discuss some features in social network messaging applications that can be used to trigger LTE paging requests to devices in which the subscriber has installed the corresponding social network applications.
%We present a new method to page LTE subscribers via their ``social identities'' in application-layer systems such as a social network messaging system or email. These identifiers can be easily acquired  over the Internet. 

%We assume that social applications (Facebook Messenger~\cite{fmessg} and Whatsapp~\cite{whatsapp}) are installed on the UE and the Internet connectivity is over LTE data connection rather than over wifi or 2G/3G networks. Now we describe methods to exploit Facebook and Whatsapp application to trigger paging messages without the user being alerted.

\begin{figure}
	\centering
  \includegraphics[width=1.4in]{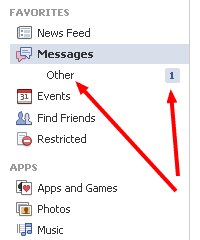}
  \caption{`Other' folder in Facebook}
  \label{fig: others folder in facebook}
\end{figure}

\noindent\textbf{Facebook `Other' message folder:}
\on{Many Facebook~\cite{facebook} users are not aware of the \emph{`Other'} message folder in Facebook (shown in Figure~\ref{fig: others folder in facebook}).}{\hypertarget{r2-4}{Many Facebook~\cite{facebook} users do not know about the \emph{`Other'} message folder (as shown in Figure~\ref{fig: others folder in facebook}) in Facebook.}}
Normally when a message is received from a Facebook friend, it will be stored in the normal inbox folder of that user.
But messages from people who are not in the friend list may be directed to the \emph{`Other'} folder. Further, the user is \emph{not notified} about messages in the \emph{`Other'} folder. In fact, the user himself has to manually check \emph{`Other'} folder to even notice that there are waiting messages.
According to Facebook~\cite{fbothers}, this is intended to protect users against spam. 
When an LTE subscriber has the Facebook application installed on his LTE device, \textit{all} incoming Facebook messages, including those that end up in the `Other' folder, trigger a paging request by the network. Other Facebook features, such as repeated friend requests or poking (depending on the user's profile settings) also trigger paging requests. However, in those cases, unlike in the case of messages that end up in the `Other' folder, Facebook application notifies the user.
%make use of this spam protection feature: Sending a message from non-friend account will cause the network to send a paging request for the subscriber, while Facebook application on the UE will not notify this action. Similarly, sending a friend request (repeated) or poking a subscriber (depending on the user's profile settings) also triggers paging message, however the user is notified in these cases.
% However, the attacker can make use of this spam protection feature: Sending a message from non-friend account will cause the network to send a paging request for the subscriber, while Facebook application on the UE will not notify this action. Similarly, sending a friend request (repeated) or poking a subscriber (depending on the user's profile settings) also triggers paging message, however the user is notified in these cases.

%Case of Twitter - \textcolor{red}{\textbf{NOT COMPLETED}} \\ 

\noindent\textbf{WhatsApp `typing notification':}
WhatsApp supports a `typing notification' feature - when someone (`sender') starts composing a message to a person (`recipient') using WhatsApp, the WhatsApp client UI at the recipient shows a notification to the recipient that an incoming message is being typed. If the recipient is using a WhatsApp client on an LTE device, this ends up triggering a paging request.

\subsection*{RRC protocol issues}

%We exploit two functions of RRC protocol information and operator configuration issues to demonstrate location leaks in practice. 
The LTE RRC protocol includes various functions needed to set up and manage over-the-air connectivity between the eNodeB and the UE as described in~\cite{36.331}. For our attacks, we exploit two of these: network broadcast information and measurement reports sent by UEs to the network.

\noindent\textbf{Broadcast information:} In this RRC protocol function, temporary identities associated with UEs (i.e., GUTIs) are transmitted over the air in a broadcast channel. Such broadcast messages are neither authenticated nor encrypted. Hence anyone can decode them with appropriate equipment. 
Since these broadcast messages are only sent in specific geographical areas, we can use the method described in~\cite{fookune2010locationgsm} to reveal the presence of subscribers in a targeted area by exploiting these broadcast messages.
%Further, broadcast messages are sent in a given tracking area or a cell, depending on previously known locations of the UE. As a result, it is possible for anyone to observe broadcast channel information in a targeted area. We use a method described in~\cite{fookune2010locationgsm} to reveal identities and the presence of subscribers in the targeted area by exploiting these broadcast messages.

Further, the eNodeB periodically broadcasts SIB messages which carry information for UEs to access the network, perform cell selection, and other information as described in~\cite{36.331}. The attacker can utilize this broadcast information to configure the rogue eNodeB for malicious purposes.

\noindent\textbf{UE measurement reports:} In LTE, UE performs network measurements and sends them to the eNodeB in RRC protocol messages when requested. Such UE measurement reports are necessary for network operators to troubleshoot signal coverage issues. In particular, there are two types of UE measurement reports  - one sent in \emph{``Measurement Report"} used as part of handover procedure and other one in Radio Link Failure (RLF) report - which are used to troubleshoot signaling coverage. However, since these messages are not protected during the RRC protocol communication, an attacker can obtain these network measurements by simply decoding from radio signals.
%without having any UE-specific keys (i.e., EPS security context).

We now explain the importance of two RRC protocol messages and measurement information they carry. First,  \emph{``Measurement Report"} message is a necessary element during handover procedure in LTE networks. Generally, eNodeB sends a RRC message indicating what kind of information is to be measured in response the UE sends \emph{``Measurement Report"} messages. We discovered that the LTE specification allows sending this message to the UE without AS security context~\cite{36.331}.
Second, RLF report is a feature to detect connection failures caused by intra-LTE mobility and inter-system handovers between  LTE, GSM, and 3G networks. Upon detection of such events, RLF reports are created by the UE and forwarded to eNodeB when requested. These reports are collected by the Operations, Administration, and Maintenance (OAM) system for troubleshooting. As per the LTE standard specification~\cite{36.331} appendix A.6, the \emph{``UEInformationResponse"} message carrying RLF report should not be sent by the UE before the activation of AS security context. However, we discovered that major LTE baseband vendors failed to implement security protection for messages carrying RLF reports. This suggests that the specification is ambiguous leading to incorrect interpretation by multiple baseband vendors.

In particular, \emph{``Measurement Report"}  and \emph{``UEInformationResponse"} messages contain serving and neighboring LTE cell identifiers with their corresponding power measurements and also similar information of GSM and 3G cells. Additionally the message can include the GPS location of the UE (and hence of the subscriber) if this feature is supported. We exploit the above vulnerabilities to obtain power measurements, which we then use to calculate a subscriber's precise location.

\subsection{Initial measurements}
  \label{sec:initialmesurements}

We performed a measurement study on LTE networks of three major operators to understand GUTI allocations, Smart Paging, and mapping of tracking area and cell dimensions for the purpose of examining the feasibility aspects of location leak attacks.

Before measuring GUTI allocations and Smart Paging, we consider the following timing constraints for the paging procedure in LTE. Paging messages are sent only if a UE is in IDLE state. During an active connection, there are no paging messages. According to~\cite{36.300}, if the UE remains silent for 10 seconds during a connection, the eNodeB releases the associated radio resources and the UE moves into IDLE state.

\noindent\textbf{GUTI variation:} GUTI reallocation depends entirely on operator configuration. We investigated GUTI allocation and reallocation methods used by several operators. Specifically, these experiments verify whether GUTIs are really temporary in practice. We used a Samsung B3740 LTE USB data stick as the UE, since it allows us to view the RRC and NAS messages in Wireshark~\cite{p1sec}. The changes in GUTI can be seen in the \emph{``Attach Accept"} or \emph{``TAU Accept"} NAS messages in the Wireshark traces. We identified these NAS messages and recorded GUTIs for every operator for further analysis. \on{} {\hypertarget{r1-9}{In addition, GUTI variation can be verified with engineering mode on few selected handsets, for example LG G3~\cite{lgg3}.}} Our results in Table~\ref{table:guti} show that GUTI allocation and reallocation mechanisms are similar among all operators. The results are summarized below:

\begin{itemize}

\item Periodically (once an hour and once in 12 hours) detaching and attaching the UE while it was stationary resulted in the same GUTI being re-allocated in all three operator networks. A stationary UE did not have its GUTI changed for up to three days or when moving between TAs within the city.

\item When UE was moving inside the city for 3 days while remaining attached to the network, no change in GUTI was observed in any operator's network. 
%When the UE was moving inside the city for 3 days and attached to the network, there is no change in GUTI allocation on all operators during \emph{``TAU Request"}. 

\item %As we observed, by completely turning off the UE for one day, a new GUTI was allocated when turned ON the following day. However, we noticed inadequate GUTI reallocation methods for a major network operator. 
If a UE was completely turned off for one day, a new GUTI was allocated when it was subsequently turned on. In the case of one of the operators, the newly assigned GUTI differed from the old one by only one hexadecimal digit. This implies that GUTIs were not chosen randomly.

\end{itemize}

Based on above observations we conclude that the GUTI tends to remain the same even if a UE is moving within a city for up to three days. Hence temporary identities are not really temporary in any of the three networks. This allows an attacker to perform passive attacks.

\begin{table}[htbp]
\small

 \centering
{ \renewcommand{\arraystretch}{1.2}
\resizebox{1.0\columnwidth}{!}{%
\begin{tabular}{ |c|c|c|c|c|c|} 

\hline
   \textbf{\textit{Activity}} &  \multicolumn{2}{c|}{\textbf{\textit{Smart Paging}}} & \textbf{\textit{GUTI changed?}} \\ [5pt] 
\cline{2-3}  
  & \textbf{\textit{on Cell}} & \textbf{\textit{on TA}}  & \textbf{\textit{(All operators)}}  \\ [3pt] \hline 
 
 \hline
 \textit{Facebook Message }& \textsl{Yes} & \textsl{No} & \textsl{No} \\ \hline 
 \textit{SMS }& \textsl{Yes} & \textsl{No} & \textsl{No }\\ \hline
\textit{ VoLTE call} & \textsl{No} & \textsl{Yes} & \textsl{No} \\ \hline
 \textit{Attach and Detach every 1 hour} & \textsl{-} & \textsl{-} & \textsl{No} \\ \hline
 \textit{Attach and Detach every 12 hour} & \textsl{-} & \textsl{-} & \textsl{No}\\ \hline
 \textit{Normal TAU procedure } &  \textsl{-} & \textsl{-} & \textsl{No} \\ \hline
 \textit{Periodic TAU procedure} & \textsl{-} & \textsl{-} & \textsl{No} \\ \hline 

%\multirow{3}{9em}{Location Leak} & passive & trade off Security and Availability & fresh GUTI reallocation  \\ [7pt] \cline{2-5} 
%& social networks & software architecture  & avoid sending silent messages to client   \\ [7pt] \cline{2-5} 
%& active & 3GPP specification and BB implementation & integrity protection   \\ [7pt] \cline{2-5} 
%\hline

%\multirow{2}{9em}{Denial of Service} & LTE networks & trade off Security and Availability & timer to recover  \\ [7pt] \cline{2-5}
%& All networks & trade off Security and Availability  & timer to recover  \\ [7pt] \cline{2-5}
%& bidding down  & 3GPP specification & integrity protection  \\ \cline{2-5}
%\hline
\end{tabular}
}
}
\bigskip
\caption{GUTI variations and Smart Paging behavior}
\label{table:guti}
\end{table}

%\begin{figure}
 % \includegraphics[width=\linewidth]{guti.png}
  %\caption{GUTI variations and Smart Paging behavior}
  %\label{fig:guti}
%\end{figure}

\noindent\textbf{Smart Paging:} We identified multiple cells in a busy TA for each operator and placed our passive LTE air-interface sniffer within each cell. The test UE was placed in one of the cells and remained stationary for the experiment duration. Table~\ref{table:guti} presents the set of activities performed to trigger paging messages. The results are summarized as follows:

\begin{itemize}

 \item Paging for Voice Over LTE (VoLTE\footnote{VoLTE stands for voice over LTE and it is for voice calls over an LTE network, rather than the 2G or 3G connections which are usually used.}) call occurs on the entire TA and paging for other IP applications occurs on the last seen cell. This is referred to as application aware paging~\cite{nokia}. Since VoLTE has higher priority and strict timing constraints compared to other data applications, the network pages the complete TA to find the UE quickly.

\item  When the UE paging is triggered via Facebook or SMS messages, sniffers detected a particular paging message only in the cell where the UE is located (or last seen). This implies that all operators are using Smart Paging.

\end{itemize}

\begin{figure}[htbp]
  \begin{center}
  \includegraphics[width=0.8\columnwidth]{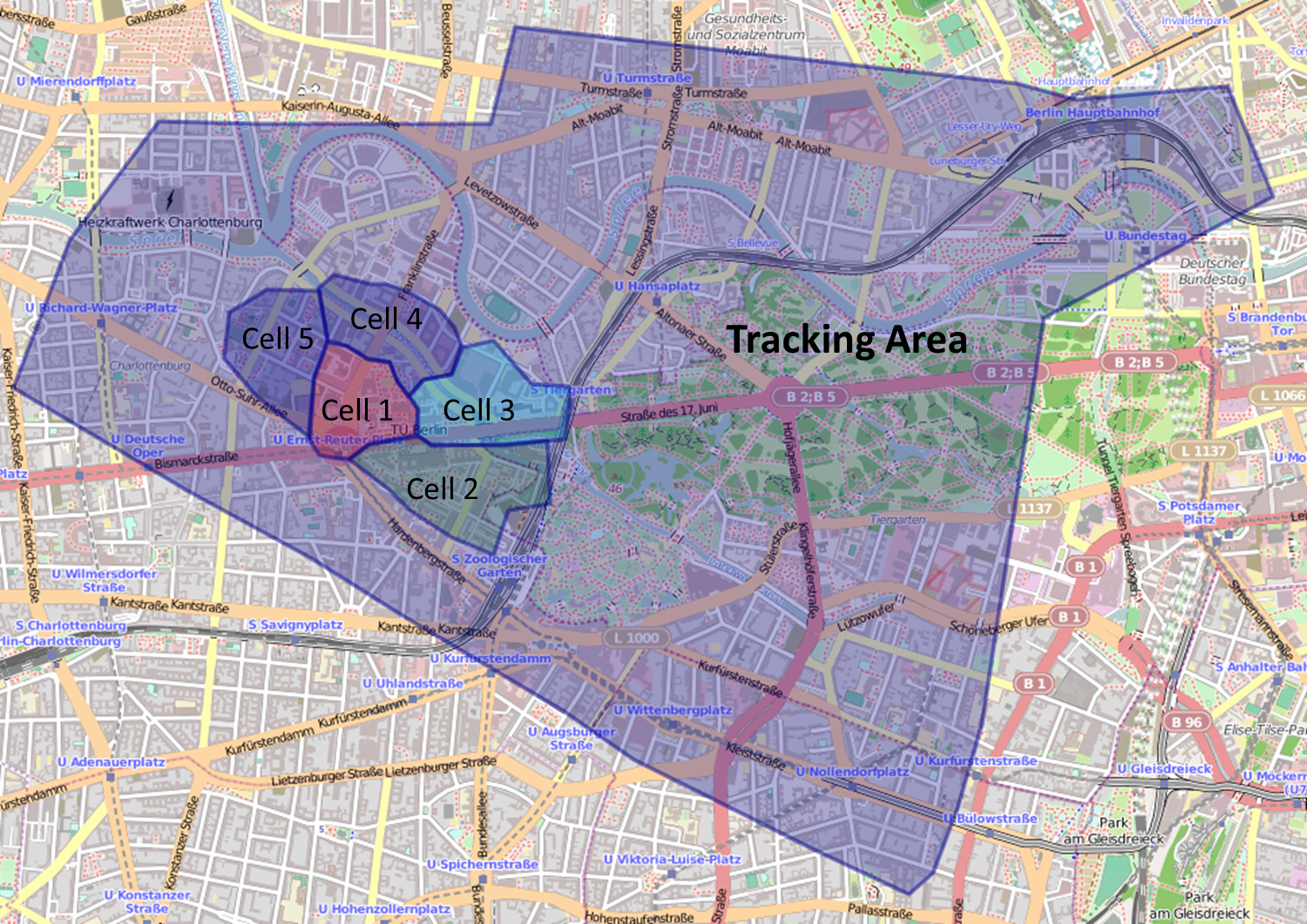}
  \caption{LTE tracking area and cells of a major operator in a city}
  \label{fig:vodafone}
  \end{center}
\end{figure}

\noindent\textbf{Mapping tracking area and cell dimensions:} It is necessary to have knowledge of the size of LTE tracking areas and cells deployed in a metropolitan city for determining a victim's location. In particular, this knowledge enables an attacker to identify targeted TAs for specific regions and network operators in the city. We created a database that maps Tracking Area Codes (TACs) to GPS coordinates by slowly bicycling through the city. The TACs are periodically broadcast in SIB Type number 1 messages~\cite{36.331}. We logged them using our passive attack setup. Further, in order to determine the surface area covered by a tracking area, we calculated the region covered  by the points with the same TAC and the results are plotted in Figure~\ref{fig:vodafone}. The size of TA inside the city varies from 10 to 30 $km^2$. According to OpenCellID~\cite{OCID} tracking areas outside the city center cover 80 - 100 $km^2$. The TAs are smaller in size compared to the GSM location areas  plotted by~\cite{nico} in the same city. 

Since the granularity we obtain through our attacks is on a cell level, it is important to know cell sizes in LTE network as compared to GSM. Further, this knowledge helps in positioning the rogue eNodeB to maximize the effect of active attacks. In order to plot cell boundaries, we used the \texttt{cellmapper}~\cite{cell-mapper} Android application which reports the cell ID, eNodeB ID, and Radio Signal Strength Indicator (RSSI) of the cell in real time. Initially, we identified a point with high signal strength (possibly close to the eNodeB) and marked it for the reference. Then we walked in all directions from the reference point till reaching the cell edge. Cell edges are identified when  RSSI becomes very poor and the UE triggers a cell change. In this way, we traced the boundaries of the 5 cells and marked them inside the TA as shown in Figure~\ref{fig:vodafone}. Based on the cell sizes measured, we find out that a major operator implemented micro cells in their LTE infrastructure. Typical size of a micro cell ranges from 200 - 2000 $m$ in radius~\cite{lte-cells}.

\subsection{Passive attack - link subscriber locations/movements over time (L1)}

In passive attack mode, attacker's objective is to collect a set of IMSIs and GUTIs which can be used for two purposes. One is to verify subscriber's presence in certain area,
%to link subscriber's presence to a certain area, 
and other is to reveal his past and future movements in that area. To achieve this, we sniff over the LTE air interface and decode broadcast paging channels to extract IMSIs and GUTIs. These identities can be collected in locations such as airports or subscriber's home or office. The attacker needs to map IMSI or GUTI associated with a particular subscriber to reveal his/her presence in that area. Since GUTI is persistent for several days in our experiments (see Section~\ref{sec:initialmesurements}), its disclosure makes the subscriber's movements linkable. The mapping between GUTI and IMSI is possible using semi-passive attacks.
%or when the network discloses IMSI-GUTI during an unexpected behavior as discussed in~\cite{33.821}. 

\subsection{Semi-Passive attack - leak coarse location (L2)}

The objective of the semi-passive attack is to determine the presence of a subscriber in a TA and further, to find the cell in which the subscriber is physically located in. In particular, we demonstrate the use of novel tracking techniques to initially determine the TA and then exploit Smart Paging to identify a cell within that TA.

%For this we implemented the method proposed %by kune et.al to reveal the mapping between the S-TMSI, the MSISDN and the social identity of the subscriber.

\subsection*{Determining tracking area and cell ID }

%As described in the section \ref{sec:initialmesurements}, we page the subscriber using VoLTE calls to determine TA and Facebook application to determine the cell.
We use following two methods to generate signaling messages for performing the attack.
 
\noindent\textbf{Using VoLTE calls}: We placed 10 VoLTE calls to the victim. The VoLTE call connection times are very short at around 3 seconds according to previous work~\cite{31}. Hence, the attacker has to choose the call duration so that it is long enough for a paging request to broadcast by the eNodeB but short enough to not trigger any notification on the UE's application user interface. As explained earlier, VoLTE has high priority and therefore its paging requests are broadcast to all eNodeBs in a TA. Hence it is sufficient to monitor any single cell within the TA for paging messages. The observed GUTIs undergo a set intersection analysis where we apply the method proposed by Kune et.al~\cite{fookune2010locationgsm} to reveal the mapping between the GUTI and phone number of the subscriber. Once successful, the presence of the subscriber is confirmed in that TA.

\noindent\textbf{Using social network and applications}: Social identities are a compelling attack vector because mobile subscribers nowadays use mobile phones for accessing popular social networks and instant messaging applications. The primary intention of the attacker is to trigger paging requests via social identities without LTE subscribers being aware of it. %In parallel, the attacker needs to map the se social identities to GUTIs.
\on{}{\hypertarget{r2-2}{For triggering paging messages, various mobile applications can be used. Due to popularity and size of user base we chose Facebook and WhatsApp applications for our experiments.}}
\on{}{\hypertarget{r1-2}{However tracking subscribers using social applications is not as effective as using VoLTE calls.}} 

We used Facebook messages as described in Section~\ref{network:issues} to trigger Smart Paging to localize the target subscriber to a specific cell. Similar to VoLTE calls, we send 10-20 messages to the subscriber via Facebook and do the set intersection analysis to link GUTIs to Facebook profiles. If the mapping is successful in a particular cell where the attacker is, the presence of the subscriber is confirmed. Otherwise the attacker needs to move to other cells and repeat the same procedure. The attacker can also place passive sniffers in every cell to speed up the localization procedure. However, this is expensive. The subscriber's presence is successfully determined in a cell a cell that is typically of size 2 $km^2$, i.e. much smaller than a GSM cell.
%of size 2 $km^2$ which is smaller than in GSM.

We also used WhatsApp similarly to exploit its ``typing notification'' feature. 
In this case, the attacker requires the phone number to identify the subscriber on WhatsApp.
In addition, the victim's privacy settings must allow the attacker to view the victim's WhatsApp profile. 
First, the attacker sends a message to the target recipient. Once it is received, the recipient's WhatsApp application will list it in the inbox.
For the attack to succeed, it is essential that the recipient does not block or delete the attacker's contact. Later, the attacker opens his active chat window corresponding to the recipient and composes a message but does not send. Due to the ``typing notification'' feature, the recipient can see that the attacker is typing in the chat window. During this procedure, network triggers paging request destined for recipient's LTE devices.

\subsection{Active attack - leak fine-grained location (L3)}
\label{active-attack}

Once the attacker determines a TA and cell where the subscriber is present, the next goal is to find his/her location more precisely. 
We now demonstrate two methods in which the attacker exploits a specification and an implementation vulnerability to this end.

\noindent\textbf{1. Via  measurement reports:} We consider a
subscriber  who is initially attached to a legitimate eNodeB. The
attacker forces him/her to attach to a rogue eNodeB by applying the
techniques mentioned in Section~\ref{buildenodeb}.
The subscriber's UE completes RRC connection procedures and initiates
a \emph{TAU} procedure with attacker's rogue eNodeB. Next, UE enters
into CONNECTED state. The attacker creates  a \emph{``RRC Connection
  Reconfiguration"} message with different cell IDs (possibly 3 or
more neighbor cells) and necessary frequencies, and sends it to the UE
without any protection. After receiving this unprotected message, UE
computes the signal power from neighboring cells and frequencies and
sends an unprotected \emph{``Measurement Report"} message to the rogue
eNodeB.

If the UE supports \emph{`locationInfo-r10'} feature~\cite{36.331}, it
includes its GPS coordinates in the measurement report. This
feature is not yet widely supported by current smartphones - however one of our test phone exhibited this behavior.   

%We consider a subscriber  who is initially attached to a legitimate eNodeB. The attacker forces him/her to attach to a rogue eNodeB by applying tricks mentioned in Section~\ref{buildenodeb}. During this step, the subscriber’s UE completes the RRC connection setup procedure and initiates a TAU procedure with attacker’s rogue eNodeB. Next, UE enters into CONNECTED state. The attacker creates  a \emph{RRC Connection Reconfiguration} message with different cell IDs (possibly 3 or more) and necessary frequencies, and sends it to the UE. After receiving this unprotected message, UE generates a measurement report and sends it to the rogue eNodeB. Specifically, these reports contain the signal power measured by the UE on the specified cells and frequencies. 

%Additionally, the attacker can create a similar message to request \emph{``locationInfo-r10"} IE feature~\cite{36.331} which force UE to add its GPS coordinates in its measurement reports. However, this feature is not supported by current smartphones as of now.

\begin{figure}[htbp]
  \begin{center}
  \includegraphics[width=0.8\linewidth]{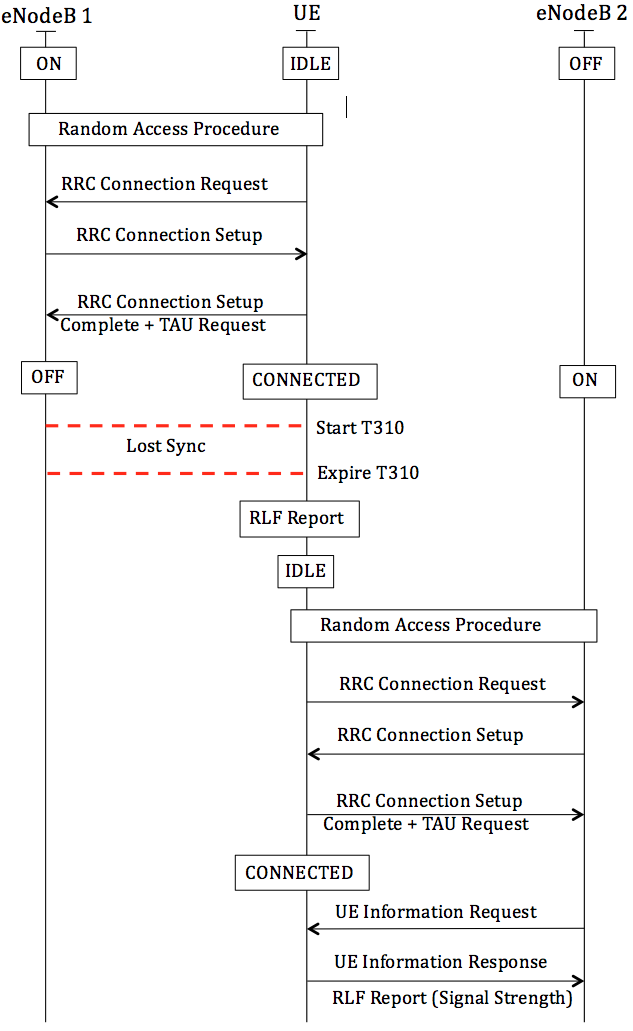}
  \caption{Retrieving RLF report from UE (L3)}
  \label{fig: Acquiring Location by creating Radio Link failure}
  \end{center}
\end{figure}

\noindent\textbf{2. Via RLF reports:}
In this attack, two rogue eNodeBs are operated in the same cell where the subscriber is present. Initially eNodeB 2 is OFF and eNodeB 1 ON to create a RLF scenario to the UE. The UE initiates connection to eNodeB 1 and enters into CONNECTED state as shown in Figure~\ref{fig: Acquiring Location by creating Radio Link failure}. We turn OFF eNodeB 1 upon receiving a TAU request from the UE. At the same time, eNodeB 2 is turned ON. Meanwhile UE detects that it has lost sync with the eNodeB 1 and starts RLF timer (T310).

When the RLF timer expires, UE creates a RLF report~\cite{36.331} and goes into IDLE mode.
In this mode, UE starts cell selection procedure as specified in~\cite{36.304} to attach to eNodeB 2. As before, UE enters the CONNECTED state with eNodeB 2 and indicates the availability of RLF report in a TAU message. Upon receiving this message, the attacker sends an unprotected \emph{``UEInformationRequest"} message to UE from eNodeB 2, thereby requesting UE to send RLF report to eNodeB 2 in response. As a result, UE sends the resulting response in an unprotected \emph{``UEInformationResponse"} message containing the RLF report. This report contains failure events and specifically signal strengths of neighboring eNodeBs.

In addition, according to the LTE specification~\cite{37.320}, RLF report can include GPS coordinates~\cite{36.331} of UE %where radio failure was experienced.
at the time it experienced the radio failure. As before, this feature is not widely implemented yet.

%\subsection{specification issue}

\subsection*{Determining subscriber's precise location}

\begin{figure}[http]
  \begin{center}
  \includegraphics[width=0.8\linewidth]{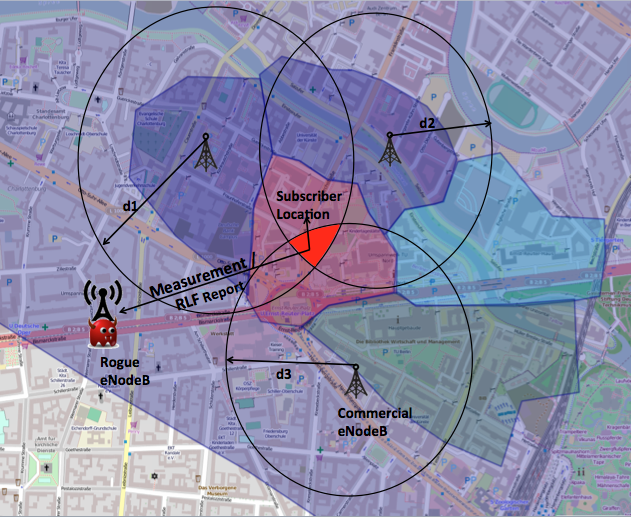}
  \caption{Determining subscriber's precise location using trilateration (L3)}
  \label{fig:locateue}
  \end{center}
\end{figure}

Aforementioned measurement and RLF reports provide signal strengths allowing the active attacker to calculate distance between the UE and the rogue eNodeB. This calculation is performed using a trilateration technique as described in~\cite{radiolocation}. Figure~\ref{fig:locateue} shows how this technique is used to determine subscriber's location. The distance estimates are calculated as d1, d2, and d3 for three neighboring base stations. The zone of intersection point of three circles is subscriber's approximate location in a cell. However, if \emph{`locationInfo-r10'} feature is supported in measurement and RLF reports, accurate location can be determined using GPS coordinates.

%TODO add a picture of locating subscbirber with having rogue eNodeB or 

 \section{DoS Attacks on LTE Air Interface}
\label{sec:dosattacks}

In this section, we demonstrate how an attacker can exploit two LTE specification vulnerabilities to deny LTE, and also GSM and 3G network services to subscribers. First, we describe the attack background and present three types of persistent DoS attacks labeled D1, D2, and D3. Later, we discuss their impact on LTE subscribers and operator services.

\subsection{Attack background}
\label{dos:background}

%We exploit the EPS Mobility Management (EMM) protocol messages which are required for control of mobility when the UE is attaching to LTE networks. In particular, we exploit two functions of EMM messages and are described as follows.
We exploit the EPS Mobility Management (EMM) protocol messages which are required for control of UE mobility in LTE networks. In particular, we exploit two functions of EMM messages described below.

%\textbf{1. LTE attach procedure:} During an attach procedure,  UE sends its capabilities list to the network. In particular, these capabilities include types of network supported (such as LTE, GSM or 3G), security algorithms, and other features as defined in~\cite{24.301}. However, these capabilities are sent in unencrypted message. To protect against MiTM attacks, the LTE security architecture mandates to reconfirm previously negotiated  security capabilities after the AKA procedure~\cite{33.401}. In particular, the network sends an integrity-protected message including the list of security algorithms previously received from the UE. There is no similar confirmation for network capabilities. Thus when an attacker removes selected network services from the list of capabilities sent by the UE, the network will accept the modified list resulting in the denial of those services to the UE.

\noindent \textbf{1. TAU procedure:} One of the main function of EMM protocol messages is to inform the network about UE's present location in the serving area of the operator. This allows the MME to offer network services to the UE, e.g., when there is a incoming call. For this purpose, UE notifies the MME of its current TA by sending a \emph{``TAU Request"} message and also includes its network modes. Generally, UE operates in various network modes for voice and data connections as stated in~\cite{24.301}, but for this work we focus only on two modes: i) EPS services (i.e., LTE services), ii) both EPS and non-EPS (i.e., GSM or 3G) services. During a \emph{TAU} procedure, the UE and MME agree on one of these modes depending on the type of subscription (for example, USIM is subscribed for LTE services), and network capabilities supported by the UE and by the operator in a particular area.

%One of the main function of EMM protocol messages is to inform the network about UE’s present location in the serving area. In LTE, the MME maintains UEs location information to find out in which TA they are present to offer network services. For this reason, UEs notify the MME of its current TA and network modes by sending a \emph{TAU Request} message. Generally, UE operates in various network modes for voice and data connections as stated in~\cite{24.301}, but for this work we focus only on two modes -  EPS services (i.e., LTE services), and both EPS and non-EPS (i.e., GSM or 3G) modes. During a TAU procedure, the UE and MME agree on one of these modes depending on the
%the type of  subscription (for example USIM is subscribed for LTE services), and network capabilities supported by the UE and operator in a particular area.

During \emph{TAU} procedure the network may deny some services to UEs, for example if the subscriber's USIM is not authorized for LTE services or if the operator does not support certain services in the serving area. The LTE specification~\cite{24.301} defines certain EMM procedures to convey such denial messages to UEs. Specifically, these are sent in \emph{``TAU Reject"} messages which are not integrity protected. 

%We identify a technical vulnerability in this specification which enables our DoS attacks. We exploit the fact that certain \emph{``TAU Reject"} messages sent from the network are accepted by UEs without any integrity protection. In particular, there is no need of mutual authentication and security contexts between the UE and network for accepting such reject messages. 
%Note that, the attacker does not require security keys specific to UEs to send \emph{``TAU Reject"} messages. Hence, the attacks can be targeted towards all LTE subscribers within the range of the rogue eNodeB. Similar types of attacks are also possible with \emph{``Service Reject/ Attach Reject"} messages.

\noindent \textbf{2. LTE Attach procedure:} During an $Attach$ procedure, UE sends a list of its capabilities to the network in an \emph{``Attach Request"} message. In particular, these capabilities include supported networks (such as LTE, GSM or 3G), security algorithms, and other features as defined in~\cite{24.301}. However, these capabilities are sent unprotected and hence, the list can be altered by an attacker. To protect against MiTM attacks, the LTE security architecture mandates reconfirmation of previously negotiated security capabilities after the AKA procedure~\cite{33.401}. In particular, the network sends an integrity-protected message including the list of supported security algorithms previously received from the UE. However, there is no similar confirmation for UE's network capabilities.

%During the TAU procedure a network may deny some of these services to the UE, for example if the subscriber's USIM is not authorized for LTE services or if the operator does not support certain services in the serving area. The 3GPP specification defines certain EMM procedures to convey such denial messages to UEs. Specifically, these are sent in \emph{TAU Reject} messages as stated in an 3GPP specification~\cite{24.301}. We identify a technical vulnerability in this specification which enables our DoS attacks. We exploit the fact that certain \emph{TAU Reject} messages sent from the network are accepted by  UEs without any integrity protection.

%In particular, there is no need of mutual authentication and security contexts between the UE and network for accepting such reject messages. Note that the attacker does not require security keys specific to UEs to send \emph{TAU Reject} messages and hence, the attacks can be targeted towards all LTE subscribers within the range of the rogue eNodeB. Similar types of attacks are also possible with \emph{service reject/attach reject} messages.

 \subsection{Downgrade to non-LTE network services (D1)}
 \label{dos1:lte}

We identify a vulnerability in the LTE specification which enables the following DoS attacks D1. We exploit the fact that certain \emph{``TAU Reject"} messages sent from the network are accepted by UEs without any integrity protection. In particular, there is no need of mutual authentication and security contexts between the UE and network for accepting such reject messages. 
Note that, the attacker does not need any security keys to send \emph{``TAU Reject"} messages. Hence, the attacks can be targeted towards any LTE subscribers within the range of the rogue eNodeB. Similar types of attacks are also possible with \emph{``Service Reject/ Attach Reject"} messages.

\begin{figure}
  \includegraphics[width=\linewidth]{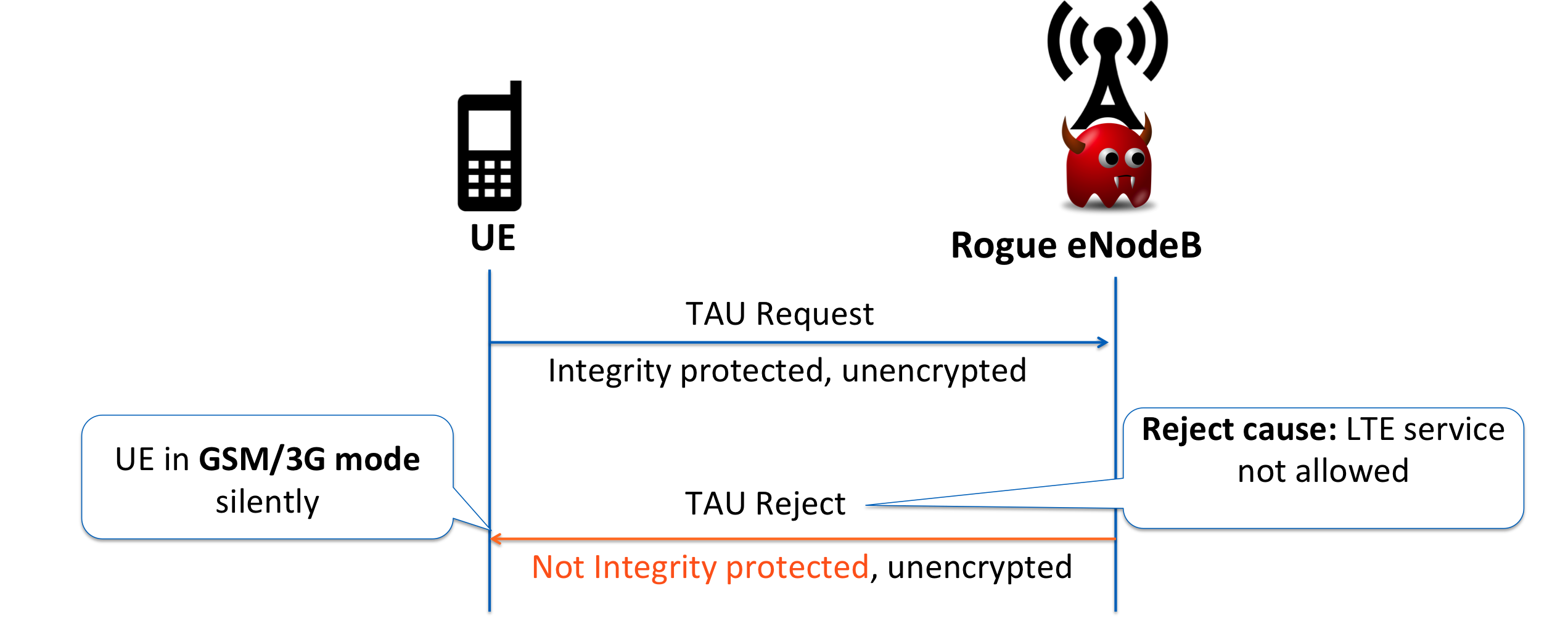}
  \caption{DoS attack - denying LTE network services (D1)}
  \label{fig:DenyingEPSservices}
\end{figure}

As shown in Figure~\ref{fig:DenyingEPSservices}, the UE sends \emph{``TAU Request"} message to attacker's rogue eNodeB. Note that as the UE is attached to the real network, this message can be integrity protected using the existing NAS security context. However, according to LTE specification~\cite{24.301}(section 4.4.5), this message is not encrypted. As a result, rogue eNodeB decodes it and responds with a \emph{``TAU Reject"} message. The attacker includes EMM cause number 7 \emph{``LTE services not allowed"} into this message. As no integrity protection is required, the victim's UE accepts the message. The UE proceeds to act on the indicated rejection cause by deleting all existing EPS contexts associated with the earlier (real) network.

As a result, UE updates its status to \emph{``EU3 ROAMING NOT ALLOWED"}\footnote{It means that last \emph{TAU} procedure was correctly performed, but reply from the MME was negative due to roaming or subscription restrictions.} and considers the USIM and hence the UE as invalid for LTE services until it is rebooted or USIM is re-inserted. Further, UE does not search for or attach to legitimate LTE networks even if they are available in that area, causing a denial of service. However, if supported, the UE searches for GSM or 3G network in the same area to gain network services. \on {} {\hypertarget{r1-6} {By downgrading subscribers, an attacker could attempt to launch known 2G or 3G attacks, besides loss of LTE services. }}
%Signaling messages exchanged between the UE and rogue eNodeB are shown in Figure~\ref{fig:DenyingEPSservices}.

\subsection{Denying all network services (D2)}
\label{dos2:lte}

\begin{figure}
  \includegraphics[width=\linewidth]{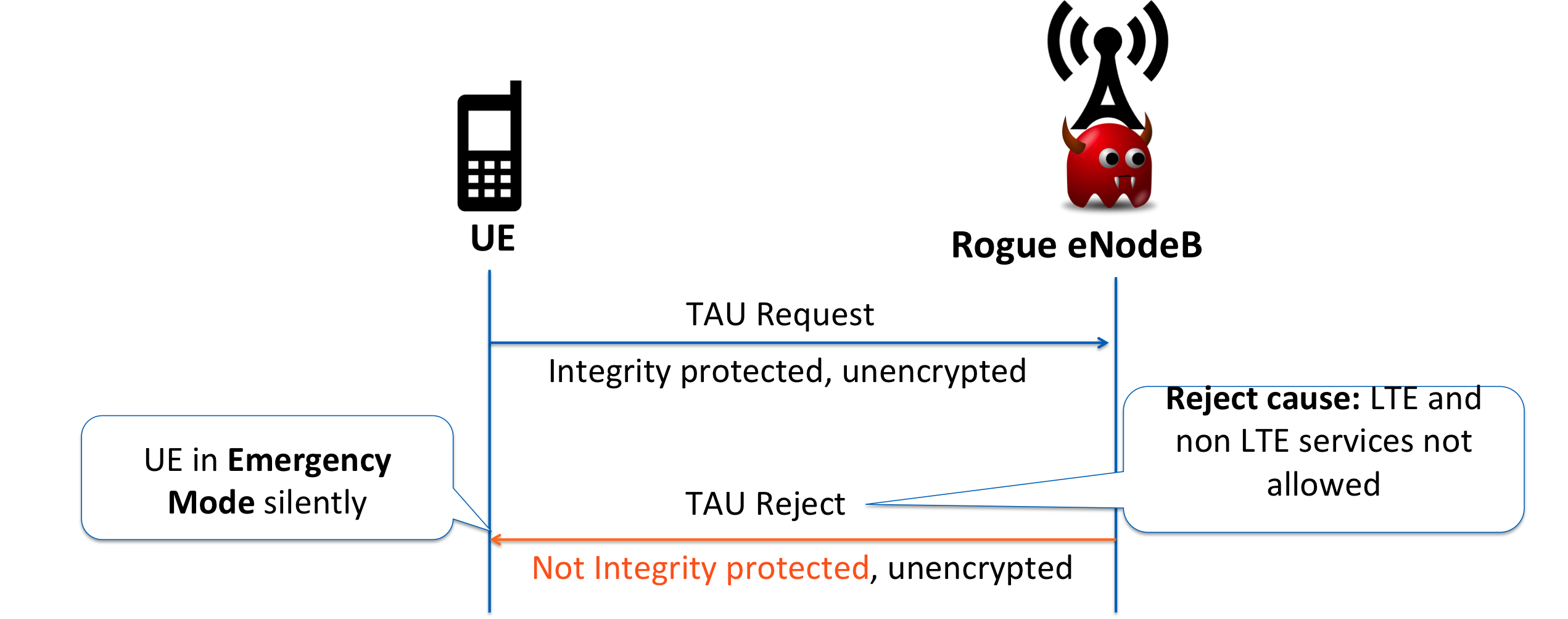}
  \caption{DoS attack - denying all mobile network services (D2)}
  \label{fig:DenyingEPSandNonEPSservices}
\end{figure}

D2 is similar to D1 but the result is different. The UE initiates TAU request procedure and rogue eNodeB responds with a TAU Reject message with the cause number 8 which is \emph{``LTE and non-LTE services not allowed"}. After receiving this message, the UE sets LTE status to \emph{``EU3 ROAMING NOT ALLOWED"} and considers USIM invalid for the network until it is rebooted or USIM is re-inserted. Further, it enters the state EMM-DEREGISTERED: UE's location is unknown to the MME and is not reachable for any mobile services. As a result, UE does not attempt to attach to LTE, GSM, or 3G networks for normal services even if networks are available. The UE remains in the EMM-DEREGISTERED state even it moves to a new TA or even to a new city, thereby causing a persistent denial of service. Signaling messages exchanged between the UE and the rogue eNodeB are shown in Figure~\ref{fig:DenyingEPSandNonEPSservices}.
%\on{}{\hypertarget{r1-3}{As discussed in Section~\ref{active-attack}, L3 attack could be a DoS attack against the attached UE, however, this DoS is not persistent.}}

\subsection{Denying selected services (D3)}
 \label{dos3:lte}

In this attack, the active attacker modifies messages exchanged between the eNodeB and UE. However, note that this attack was not performed during our experiments due to unavailability of UE baseband software.

The UE initiates an \emph{``Attach Request"} message to the eNodeB and this message is intercepted by the attacker. The message contains \emph{``Voice domain preference and UE's usage setting"}  which informs the network about UE's voice calling capabilities. The attacker removes these capabilities from this unprotected message and adds \emph{``Additional update type - SMS only"} before forwarding it to the network.
The network accepts this message and executes AKA protocol with the UE to complete $Attach$ procedure. 
However at this step, the MME configures UE's profile with the received (modified) capabilities, thereby allowing only SMS and data services.  When there is an incoming call for UE, the MME rejects it and informs the cause to the subscriber who is calling.  On the other hand, if UE tries to make an outgoing voice call, the network rejects this request and informs the cause. This is an example of a bidding down attack. The denial is persistent since the attack is effective even after the attacker has moved away. However, the user can recover from the attack by restarting the UE or moving to another TA. 3GPP specifications does indeed mention a timer (T3245) that a UE can use to recover from EMM DISCONNECTED state~\cite{24.008}. However, the use of this timer is optional (none of the devices we tested implement this timer). The default timer value (24-48 hours) is too large in the case of DoS attacks.
 
%Note that the victim may be able to serve voice calls to the UE via VoLTE services if supported by the operator.
%@electronic{voicecalls,
%  author    	= "GSMA",
%  title     	= "VoLTE Service Description and Implementation Guidelines",
%  url       	= "http://www.gsma.com/network2020/wp-content/uploads/2014/10/FCM.01-VoLTE-Service-Description-and-Implementation-Guidelines-Version-2.0.pdf",
%}

\iffalse

	\begin{savenotes}
		\begin{table}
		    \caption{List of tested LTE phones, baseband vendors and affected vulnerabilities }
	\label{table:test}
	\begin{tabular}{|p{1.5cm}|l|l|l|l|l|}
	    \hline
	    \multirow{2}{*}{\parbox{1.5cm}{Phone Model}}
	    & \multirow{2}{*}{\parbox{1.5cm}{Baseband Vendor}}  & \multicolumn{2}{|c|}{Affected Vulnerabilities} & \multirow{2}{*}{\parbox{1.7cm}{DoS Recovery}} \\ [0.2cm]
	    \cline{3-4}
	    &  & DoS & Location Leak &  \\ [0.1cm]
	    \hline
	    \hline
		iPhone 6 & Qualcomm & Yes & Yes & r\footnote{reboot},re\footnote{re-insert USIM} \\[1ex]
		\hline
		 Samsung S6 & Samsung & Yes & Yes & r,re,f\footnote{flight mode toggle} \\ [1ex]
		 	\hline
		Lenovo K50  &  MediaTek & Yes & Yes  & r,re,f \\[1ex] \hline
		Huawei P8  & Huawei & Yes & Yes & r,re,f \\ [1ex] \hline
	\end{tabular}
	\end{table}

	\end{savenotes}

\fi

 \subsection{Impact on end-users and operators}

%Generally DoS attacks are threat to availability and reliability of network threats

%Unlike the LTE jamming DoS attacks described in~\cite{ltedosattacks}, our attacks are against UEs in a certain area instead of against the LTE networks. 
%A successful attack would deny the target UE from utilizing mobile services temporarily. Typically, the UE remains in non-service state for this time period even if the attacker shuts down his rogue eNodeB or moved away from attacking area. Because the effect of the attack persists even after the active attacker is not present, this attack is more significant than other types of non-persistent DoS attacks (for example jamming and RACH flood~\cite{rach} that are difficult to prevent). The present LTE security architecture could not prevent this type of DoS attacks. Impact of DoS attacks are as follows:

Unlike the LTE jamming DoS attacks described in~\cite{ltedosattacks}, our attacks are against UEs in a 
certain area instead of against LTE networks. A successful attack would deny the target UE from utilizing network services. 
Typically, the UE remains in non-service state for some time period even if the attacker shuts down his rogue 
eNodeB or moves away from attacking area. 
Consequently, this attack is more serious than other types of DoS attacks 
(for example jamming and RACH flood~\cite{rach} that are difficult to prevent). Impact of these attacks are as follows:
%Further attacks are of persistent types since UEs are affected even if the adversary is moved away from attacking area. 

\begin{itemize}

	  \item Subscriber's UE may not alert the user about the unavailability of legitimate services. 
	  However, depending on the alert notification capabilities provided by application layer of 
	  various mobile operating systems installed on the UE, the subscriber could be notified of limited services or no network connectivity status. 
 	  \on{}{\hypertarget{r2-3}{We noticed that there is no standard approach across different mobile operating systems to indicate the type of active network mode (e.g., 2G/GSM, 3G, LTE) to the user.}}

 \item \on{Subscribers will not be able to receive or make any calls and data connections.}{\hypertarget{r2-9}{Subscribers will not be able to receive or make normal calls and data connections.}}
 Hence, a significant loss is incurred to both network operators and their subscribers.
  Network operators are not able to offer services since subscribers are unavailable technically and no billing would occur.

\item UE can still make emergency calls. However, emergency calls are not possible when UE is attached to a rogue eNodeB.

\item LTE-capable M2M devices which are not attended by technicians on a daily basis could be blocked 
out from network services for a long time. This is due to the fact that M2M devices need to be rebooted or USIM needs 
to re-inserted to recover from the attacks. 

	%M2M devices that function unattended for several years c blocked out from using mobile services.

\end{itemize}

 \section{Attack Feasibility and Amplification}
\label{sec:feasi}

 %All the LTE capable UE's do support of tracking area update reject messages and the emm reject causes without any security protection. Hence all smartphones with LTE support are vulnerable to the described DoS attacks. The impact of the attack is huge and only needs an exchange of only 2 NAS messages between the UE and the rogue eNodeB. 

%The vulnerability we exploited to mount successful DoS attacks is in the LTE specification rather than in the UE's software. Therefore, all LTE-capable UEs that implement mandatory requirements as per the standard are affected. With the rogue eNodeB setup, we can deny mobile services to UEs in the radius of 50 to 100 meters. 

%On the UE side, the LTE protocol stack is implemented in the baseband software. Primarily there are four vendors  providing LTE baseband software: Qualcomm, MediaTek, Samsung, and Huawei. We tested a few phones from every LTE baseband vendor as listed in Table~\ref{table:test} to investigate how they recover from DoS attacks. As shown in the table, UEs having a Qualcomm baseband recover after reboot or re-inserting USIM whereas other phones recover after flight mode toggle. In addition to these recovery methods, it is sufficient to toggle the flight mode feature to regain the network services for UE's having Samsung, Mediatek and Huawei baseband. 
%A reboot of the UE or the USIM re-insertion is required in the case of UE's having a Qualcomm baseband.

%Note that, the Table~\ref{table:test} does not contain experimental results for denying voice call services DoS attack described in Section~\ref{dos3:lte}.

In this section, we discuss the feasibility of both location leak and DoS attacks against popular LTE smartphones and methods to amplify the coverage range of our attacks.

Several of the vulnerabilities we exploited are in the LTE specifications rather than in the UE's baseband software.
 %All these specifications and related protocols are implemented in the UE's baseband software. 
Therefore, all LTE-capable UEs conforming to these specifications are affected. 
For evaluation, we selected popular smartphones incorporating baseband implementations from top vendors who dominate the market share worldwide~\cite{idc}. 
We successfully verified that all these phones are vulnerable to our attacks.
In addition, all UEs have the implementation vulnerability leading to attack L3. 

%Although this vulnerability is due to implementation error in baseband software but indicates ambiguity in implementing 3GPP standards by all baseband vendors. 

We further investigated on how UEs recover from DoS attacks. We found out that all UEs recover after rebooting or re-inserting the USIM. Additionally, UEs having baseband from most vendors can recover by toggling the flight mode.

\noindent \textbf{Attack amplification:} Related to our passive attacks, we determined the average cell radius of a major operator in a city is 800 meters for the 2.6 GHz and 1 km for the 800 MHz frequency band. The USRP B210 used for our attacks has a maximum output power of 20dbm (100mW)~\cite{usrp-power} with a coverage range of 50 to 100 meters. However, the signal coverage  area can be increased with a suitable power amplifier. 
Specifically, based on the COST 231 radio propagation model~\cite{cost123}, we calculated that by mounting a USRP at a height of 10m (e.g., on a street lamp) and amplifying the power by 10 dB, it is possible to deny LTE and non-LTE services for every subscriber in a cell. 
For a reference, \texttt{OpenBTS} projects~\cite{openbtsrange,umtrx} use USRPs to provide GSM coverage in rural areas with $>$2 km coverage with an external power amplifier and antenna. Similarly, signal coverage area of our rogue eNodeB could be increased to demonstrate feasibility of the attack.

%In the reverse or uplink the rogue eNodeB should also have high receiver sensitivity given that most LTE handsets are able to transmit at 23dBm (0.2 W)~\cite{36.101}.
 %Based on our passive attacks, we determined the average cell radius for Vodafone in Berlin is 1 km for 2.6 GHz and 3 km for 800 MHz in dense urban areas. Using the COST 231 radio propagation model~\cite{cost123}, we calculated that by mounting a USRP at a height of 30m and amplifying the power by 40 dB it is possible to deny LTE and non-LTE services for every LTE subscriber inside a cell. As defined in~\cite{36.942} the typical power of an eNodeB in the macro cell in an urban area can be upto 46 dbm (40 watts). The USRP B210 used for the attack has a maximum output power of 20dbm (100mW)~\cite{usrp-power} but this can be amplified with a suitable power amplifier. In reality,  the OpenBTS projects~\cite{openbtsrange,umtrx} use USRP's to provide GSM coverage in rural areas with $>$2km coverage with an external power amplifier and antenna. In the reverse or uplink the rogue eNodeB should also have ample receive signal power given that most LTE handsets are able to transmit at 23dBm (0.2 W)~\cite{36.101}. 
 
 \section{Security Analysis}
\label{sec:securityanalysis}

\on{In this section, we discuss vulnerabilities discovered in this paper and their impact on LTE security.}{\hypertarget{r2-8}{In this section, we discuss vulnerabilities discovered in the specifications and their impact on LTE security}}. We explain the background behind the vulnerabilities by considering various trade-offs between security and criteria like availability and performance. We show that the equilibrium points in the trade-offs have shifted today compared to where they were when the LTE security architecture was being designed. We also discuss countermeasures for the vulnerabilities that made our attacks possible. Table~\ref{tab:security} summarizes our analysis.
%These findings also identify weaknesses in access network protocols with respect to important LTE security aspects. Table~\ref{tab:security} summarizes our analysis.

\subsection{Possible trade-offs}
\noindent\textbf{Security vs Availability:} We demonstrated a vulnerability in the LTE RRC protocol specification that allows the adversary to obtain unprotected measurement reports from UEs (L3). We consider the following two angles to explain the trade-off. 
On one hand, in some cases network operators require unprotected reports for troubleshooting purposes. In particular, if the UE is not able to establish connection with the eNodeB then it may be necessary to send measurement reports without protection in order allow the network to identify technical reason behind the fault. This seems to be the reasons behind the note in LTE RRC specification which points out that the 3GPP Radio Access Network (RAN2) working group decided to permit UEs to send reports even without security activation~\cite{36.331}.  
On the other hand, during the design work for the LTE security architecture, the 3GPP security working group (SA3) suggested that all RRC protocol messages should be sent in encrypted form~\cite{33.821}. Hence, the vulnerability in RRC protocol specification is a conscious exception to this security design guidance~\cite{36.331}. Clearly, 3GPP has concluded that in this particular case the requirement of having network availability all the time to all UEs outweighs security concerns related to subscribers' privacy.

%We demonstrated a vulnerability in the LTE RRC protocol specification that allows the adversary to obtain measurement reports from UEs without any protection. We consider the following two aspects to explain the trade-off. First,  in some cases 
%network operators require unprotected reports for troubleshooting purposes. In particular,  if the UE is not able to establish connection with the eNodeB then it may be necessary to send these measurement reports without protection in order to identify technical reason behind the fault. This seems to be the reason behind the note in LTE RRC specification which points out that Radio Access Network (RAN2) group agreed to send reports without security activation~\cite{36.331}.  
%Second, during the design work for the LTE security architecture, the 3GPP security group suggested that measurement reports should be sent in encrypted form~\cite{33.821}. However, the vulnerability in RRC protocol specification contradicts with this security design choice. These two aspects show a trade-off between security and availability requirements. Clearly, in this case the requirement of having network availability all the time to all UEs outweighs security concerns related to subscribers' privacy.

%On the other hand during an \emph{Attach} procedure, UE’s security capabilities are replayed but not the network capabilities to protect against the bidding down attack. This vulnerability as discussed in~\ref{dos3:lte} indicates that security is preferred over the availability which deny call services to UEs.

\noindent\textbf{Security vs Performance:} We observed that UEs are required to reboot or re-insert USIM after DoS attacks in order to regain network services. This behavior, exhibited by all LTE devices we tested,  is according to the LTE specification. Since the network denies services for valid reject causes described in~\cite{24.301}, the UE restricts itself from re-initiating LTE (or any mobile network) \emph{Attach} procedure in order to conserve battery power. In addition, frequent unsuccessful \emph{Attach} requests from UEs would increase signaling load on the network. These are the reasons why the LTE specification requires the UE to reboot or re-insert USIM to recover from reject messages. This preference of performance 
over security leaves LTE subscribers vulnerable to the DoS attacks (D1 \& D2).

As another example, during \emph{Attach}, UE's security capabilities are sent back to it for confirmation after security activation in order to protect against bidding down attacks. This is an application of the well-known `matching history' principle used in security protocol design~\cite{extendprotocolmatch}. However, UE's network capabilities are not protected in similar manner, enabling a different type of bidding down attack (D3). The reason for not applying the matching history principle to all negotiated parameters, as discussed in~\ref{dos:background}, indicates another trade-off where added security has not outweighed performance loss due to the full application of the matching history principle. To apply the matching history principle to all parameters would have required the inclusion of a cryptographic hash of all the parameters, instead of the parameters themselves. However, confirming only the security information capabilities, which take up much less space (only a few bits) compared to a full cryptographic hash, minimizes the overhead in signaling. 

A third example we observed is that in some operator networks, GUTIs are not changed even after three days of usage (L1). LTE specifications do not mandate any GUTI reallocation frequency, leaving it to as a policy decision to operators. One possible reason for the low GUTI-change frequency is the operators' wish to reduce signaling overhead by trading off privacy.

\noindent\textbf{Security vs Functionality}: Our attacks that leak coarse-grained location information by using social network messaging services (L2) is an example of the tension between security and functionality. The introduction of TCP/IP based data communication on top of mobile communication infrastructures has greatly expanded the functionality that third party developers can build for these networks. But such a flexible software architecture makes it harder to avoid or detect the type of vulnerability that led to this attack. Furthermore, even if individual app developers would fix their applications (e.g., Facebook could change the application architecture of their Messenger application to ensure that messages that end up in the "Other" box do not trigger paging requests), other application developers may make similar mistakes. To avoid such vulnerabilities in a modern mobile communication system like LTE, it would require significant developer outreach and education to help them design and build mobile optimized applications~\cite{appgsma}.

\noindent\textbf{Summary}: The design philosophy of LTE security required leaving some safety margin in security mechanisms in order to protect against changes in trade-offs. However, in the above cases the safety margins turn out to be too narrow. As a general learning on an abstract concept level, it would be better to include agility in the security mechanisms instead of a rigid safety margin. The forthcoming fifth generation (5G) technology will offer better possibilities to engineer agility and flexibility for security because software defined networking and cloud computing are among the key concepts of emerging 5G architectures.

3GPP follows the good practice of documenting exceptions when  specification needs to deviate from the general security design principles recommended by the security working group (as was the case with L3 or D1/D2/D3). We recommend further that each such exception should also trigger an analysis of its implications. For example, if an exception is made to forego integrity protection for a denial message from the network, then the standards group should consider what happens and how to recover if the denial message contains incorrect information.

\begin{table*} 
%\footnotesize

 \centering
{\renewcommand{\arraystretch}{1.7}
\resizebox{1.0\linewidth}{!}{%
\begin{tabular}{ |c|c|c|c|c|c|} 

\hline
  \multicolumn{2}{|c|}{\textbf{\textsf{Attack}}} & \textbf{\textsf{Adversary }}  &  \multicolumn{2}{c|}{\textbf{\textsf{Vulnerability}}} & \textbf{\textsf{Potential fix}} \\ [7pt] 
\cline{1-2} \cline{4-5} 
 \textbf{\textsf{Group}} & \textbf{\textsf{Description}} & \textbf{\textsf{Type}} &  \textbf{\textsf{Type}} & \textbf{\textsf{Possible trade-off}} &  \\ [2pt] \hline 
 \multirow{3}{5em}{\emph{\textsf{Location Leak}}} &\textsf{ Link location over time (L1)} & \textsf{Passive} & \textsf{Underspecification }& \textsf{(Perceived) security vs availability} & \textsf{Policy to guarantee GUTI freshness} \\ [3pt] \cline{2-6} 
 
 & \textsf{Leak coarse-grained location (L2)} & \textsf{Semi-passive}  &  \textsf{Application software architecture} & \textsf{Security vs functionality}& \textsf{Tools like Darshak~\cite{darshak} \& \on{} {\hypertarget{r1-10} {SnoopSnitch~\cite{snoop}}} to visualize suspicious signaling to subscribers}  \\ [3pt] \cline{2-6} 
 
 & \textsf{Leak fine-grained location (L3)} & \textsf{Active} & \textsf{Specification \& implementation flaw} & \textsf{(Perceived) security vs availability } & \textsf{Network authentication for requests; ciphering for responses} \\ [3pt] \cline{2-6} 
 \hline
 
 \multirow{3}{5em}{\emph{\textsf{Denial of Service}}} & \textsf{Downgrade to non-LTE services (D1)}& \textsf{Active} & \textsf{Specification flaw }& \textsf{Security vs performance} & \textsf{Timer-based recovery} \\ [3pt] \cline{2-6} 
 & \textsf{Deny all services (D2)} & \textsf{Active } &  \textsf{Specification flaw} & \textsf{Security vs performance} & \textsf{Timer-based recovery }\\ [3pt] \cline{2-6} 
 & \textsf{Deny selected services(D3)} & \textsf{Active }& \textsf{Specification flaw }& \textsf{Security vs performance} & \textsf{Extend ``matching conversation" check to more (all) negotiation parameters} \\ [3pt] \cline{2-6}
 \hline
%\multirow{3}{9em}{Location Leak} & passive & trade off Security and Availability & fresh GUTI reallocation  \\ [7pt] \cline{2-5} 
%& social networks & software architecture  & avoid sending silent messages to client   \\ [7pt] \cline{2-5} 
%& active & 3GPP specification and BB implementation & integrity protection   \\ [7pt] \cline{2-5} 
%\hline

%\multirow{2}{9em}{Denial of Service} & LTE networks & trade off Security and Availability & timer to recover  \\ [7pt] \cline{2-5}
%& All networks & trade off Security and Availability  & timer to recover  \\ [7pt] \cline{2-5}
%& bidding down  & 3GPP specification & integrity protection  \\ \cline{2-5}
%\hline
\end{tabular}}}

\bigskip
\caption{LTE attacks, vulnerability analysis, and fixes }
\label{tab:security}
\end{table*}

\subsection{Countermeasures and discussion}
We now discuss potential countermeasures against attacks demonstrated in earlier sections. In particular, we identify protocol-level and operational fixes that can be implemented by baseband vendors and mobile network operators. Some of these countermeasures are much more straight-forward than others. Similarly, some of our proposals may cause hidden dependencies and more changes may be needed in the networks than what is apparent from our descriptions.   

\noindent\textbf{Protection against location leaks:} LTE broadcast information include subscriber identities which enable tracking of UEs (L1 and L2). The broadcast information must be sent in unprotected messages from LTE system design perspective. There are two solutions to avoid UEs being tracked. One solution is to protect broadcast messages using a public key mechanism but this requires relatively big changes in LTE protocols. According to~\cite{fforsberg2012lte}, 3GPP decided against usage of public key mechanisms %due to risk of disabling legitimate subscribers and 
because its implementation cost was deemed too high. However, our findings may have changed the equilibrium in this trade-off. Consequently, a scheme where public/private keys are used only for network elements could possibly be justified now. Messages from the network could be signed by using a public key digital signature mechanism; UEs would then be able to verify the authenticity of such messages. This would prevent rogue network elements from sending false information, e.g., false messages indicating radio link failures (L3). Messages towards the network could be encrypted using the public key of the serving operator; UEs would not need to send their identities in the clear to initiate network \emph{Attach} procedure. It is not easy to protect paging messages with public key mechanisms, even if we would have public keys for UEs because UEs would have to try to decrypt all paging messages. All these proposed fixes require 
% secure way of sharing keys and 
ensuring global availability and verifiability of public keys of network components (such as eNodeB).

The second solution is more realistic as it does not require change in protocols. Network operators would simply re-allocate GUTIs often enough to avoid tracking. 
% It would still protect against passive adversaries. 
One of the national operators to whom we reported our findings, acknowledged the feasibility of our attacks and already configured their networks to prevent tracking based on GUTIs. This solution would protect against passive attacks (L1). A certain degree of protection against semi-passive adversaries could be achieved by making the adversary's actions more visible to the subscriber. There are already such tools~\cite{darshak,snoop} available but the challenge is in making them usable and useful to all types of subscribers. LTE specification vulnerability regarding UEs sending measurement reports without integrity protection needs to be addressed by the 3GPP security group in order for all baseband vendors to eventually implement the fix in their products. The simplest solution is to transmit measurement reports only after setting up the security context.
\noindent\textbf{Protection against DoS:}
The specification vulnerabilities responsible for DoS attacks based on
\emph{TAU} procedure (D1 and D2) can be fixed without changes in the
protocol itself. The 3GPP SA3 group may propose a new mechanism based
on a counter or timer value to recover from DoS attacks. If the UE is
detached from the network for a certain duration as a result of a \emph{TAU} reject messages, it should reset the configuration settings in the USIM or baseband to re-attach itself with the network without bothering the user, i.e., without having to reboot or require re-insertion of USIM. If there is an infrastructure to support distribution of operator public keys, \emph{TAU} reject messages could be signed by the network and verified by UEs. 

Next, we discuss protection against DoS stemming from bidding down attacks (D3). 
During an \emph{Attach} procedure, the UE's network and security capabilities are sent to the network.  The attacker can modify this list to downgrade capabilities reported by the UE and and forward it to the network.
To protect against such modification, both 3G and LTE contain the partial `matching history' mechanism discussed above. This allows UE to check that its original list of security capabilities are identical with the ones received by the network.  
%However, UE's network capabilities are not protected in similar manner. If the attacker changes these network capabilities, there is no way for the UE and network to verify them afterwards. 
We argue that similar protection for network capabilities is required due to the fact that the DoS attack has a persistent nature. This would of course require change in the LTE protocols. 
Again, with the use of operator public keys, it would be possible to use digital signatures to protect lists of capabilities broadcast by the network.
Alternatively, the negotiation of network capabilities could be done after AKA is successfully completed.

 \section{Related work}
\label{sec:relwork}

In this section, we describe related work in GSM, 3G, and LTE air-interface security area. \on {}{ \hypertarget{r4-4}{Previous works have reported attacks against 2G and 3G access network protocols~\cite{fookune2010locationgsm,4215735}, core network protocols~\cite{DBLP:conf/ndss/GoldeRB12,Traynor:2007:ACI:1362903.1362924,DBLP:conf/ndss/RacicMCL08,DBLP:conf/ccs/TraynorLORJMP09}, as well as services~\cite{DBLP:conf/ccs/EnckTMP05}}}. In passive attacks, Kune et al.~\cite{fookune2010locationgsm} showed that despite the use of temporary IDs, the location of a subscriber's UE in a GSM network can be leaked. In specific, it was shown that an attacker can check if a UE is within a small area, or absent from a large area, without subscriber's awareness. However, their location leaks granularity is lower and it is improved with our attacks on LTE networks. The 3GPP discuss a set of threats exposed in E-UTRAN~\cite{33.821} during LTE security study. However, the attacks we presented are not identified by the study. In active attacks, the authors in~\cite{Arapinis:2012:NPI:2382196.2382221} present a method to determine the presence of a subscriber in a particular area by exploiting a vulnerability in 3G AKA protocol. By leveraging a rogue eNodeB (femtocell), previously captured authentication parameters are replayed to the UE and the presence is confirmed based on the response from the phone. However their attack cannot reveal approximate location of the UE in a given area.

In DoS attacks, the authors in~\cite{ltedosattacks} describe that unauthenticated attach requests sent from a compromised UE/eNodeB to flood the MME and in turn to the HSS, leading to a DoS attack. However their DoS attacks are against the network and not against LTE subscribers. Through simulations the authors in~\cite{DoSLTERAN} show that Botnets can cause DoS attacks by exhausting subscriber traffic capacity over the air interface. A proof of concept paper by P. Jover et al.~\cite{rpiqjoverjam} provides an overview of new effective attacks (smart jamming) that extend the range and effectiveness of basic radio jamming. However according to~\cite{fforsberg2012lte}, both aforementioned flooding and jamming attacks are non-persistent DOS attacks hence not considered as a threat to address in the LTE architecture. In contrast, our DoS attacks are persistent and targeted towards the UE (subscribers). LTE security architecture and a detailed list of security vulnerabilities existing in the LTE networks have been presented in ~\cite{ltesecthreatsurvey}. Our attacks are not presented in this survey. Two recent papers~\cite{DBLP:conf/ccs/KimKKHJHKK15,DBLP:conf/ccs/LiTPYLLW15} discuss resource stealing and DoS attacks against VoLTE, whereas our focus is against LTE access network protocols. To the best of our knowledge, there was no previous work evaluating practical attacks on LTE access networks in the literature. 

 \section{Conclusion}
\label{sec:conclusion}

%The LTE access network security protocols promise several layers of protection techniques to prevent tracking of subscribers and ensure availability of network services at all times. 
We have shown that the vulnerabilities we discovered in LTE access network protocols lead to new privacy and availability threats to LTE subscribers. We demonstrated that our attacks can be mounted using open source LTE software stack and readily available hardware at low cost.
The need for engineering the correct trade-offs between security and other requirements (availability, performance, and functionality) led to the vulnerabilities in the first place. Such trade-offs are essential for the success of any large-scale system. But the trade-off equilibrium points are not static. We recommend that future standardization efforts take this into account. 
%We tested several handsets with LTE support of major baseband vendors and demonstrated that all of them are vulnerable to our attacks. In addition, we demonstrated our new privacy attacks in real LTE networks of several major operators. We also showed how these privacy threats can be exploited by using social identities (Facebook and Whatsapp) of subscribers. 

\on{} {\hypertarget{r1-5} {
\noindent\textbf{Impact}: We followed standard responsible disclosure practices of all affected manufacturers. We also notified affected operators as well as the standards body (3GPP). All four manufacturers acknowledged our report. Two of them have already released patches~\cite{huawei1,qual}. Two of three operators have fixed the configuration issues in their networks. 3GPP has initiated several updates to the LTE specifications to address the issues we raised~\cite{sa3meet}.
Up-to-date information about impact may be found in the arXiv report version of this paper~\cite{fullpaper} and on our project website\footnote{\url{http://se-sy.org/projects/netsec/lte/}}.}}

%The research results exhibit need ..... 


% Acknowledgements in the non-anonymous version
%\iffullversion
%\else
\noindent\textbf{Acknowledgments}: 
This work was supported in part by the Intel Collaborative Research
Institute for Secure Computing\footnote{\url{http://www.icri-sc.org/}}, Academy of Finland (``Cloud Security Services" project \#283135), Deutsche Telekom Innovation Laboratories (T-Labs)\footnote{\url{http://www.laboratories.telekom.com/public/english/}} and 5G-Ensure (grant agreement No. 671562)\footnote{\url{www.5Gensure.eu}}. T-Labs, Aalto University, and Huawei provided test devices used in our
experiments.  We thank Stefan Schr\"oder, Peter Howard, Steve Babbage, G\"unther Horn,
Alf Zugenmeier, Silke Holtmanns,
and the anonymous reviewers for their thoughtful feedback on previous
versions of this paper.

%\fi



%\bibliographystyle{IEEEtranS}
\bibliographystyle{IEEEtran}
\bibliography{lte-euro} 
\FloatBarrier

\begin{table}
\small
	
\begin{tabular}{ |l|l| }
  \hline
  \multicolumn{2}{|c|}{\textbf{Acronyms}} \\
  \hline
  3GPP & Third Generation Partnership Project \\  
  AKA & Authentication and Key Agreement \\
  AS & Access Stratum \\
  DoS & Denial-of-Service \\
  E-UTRAN & Evolved Universal Terrestrial Radio Access Network \\
  EMM &  EPS mobility Management \\
  eNodeB &  evolved NodeB \\
  EPC & Evolved Packet Core \\
  EPS & Evolved Packet System \\
  GPS & Global Positioning System \\
  GSM & Global System for Mobile Communication \\
  GSMA & GSM Association \\
  GUTI &  Globally Unique Temporary Identifier \\
  IMSI &  International Mobile Subscriber Identifier \\
  LTE & Long Term Evolution \\
  M2M & Machine to Machine \\
  MME & Mobility Management Entity \\
  NAS &  Non-Access Stratum \\
  RLF & Radio Link Failure \\
  RRC & Radio Resource Control \\
  RSSI & Radio Signal Strength Indicator \\
  SIB & System Information Block \\
  TA & Tracking Area \\
  TAC & Tracking Area Code \\
  TAU & Tracking Area Update \\
  UE & User Equipment \\
  UMTS & Universal Mobile Telecommunication System \\
  USIM & Universal Subscriber Identity Module \\
  USRP & Universal Software Radio Peripheral \\
  VoLTE & Voice over LTE \\
  \hline
\end{tabular}

\caption{Summary of Acronyms\label{tab:abbrev}}
\end{table}

\end{document}